\documentclass[10pt,journal,twocolumn]{IEEEtran}

\usepackage{epsfig}
\usepackage{times}
\usepackage{float}
\usepackage{afterpage}
\usepackage{amsmath}
\usepackage{amstext}
\usepackage{amssymb,bm}
\usepackage{latexsym}
\usepackage{color}
\usepackage{graphicx}
\usepackage{amsmath}
\usepackage{amsthm}
\usepackage{graphicx}
\usepackage[center]{caption}
\usepackage{pstricks}
\usepackage{subfigure}
\usepackage{booktabs}

\newtheorem{theorem}{Theorem}

\newcommand{\gap}{8.5048}
\newcommand{\gapz}{4.507}
\newcommand{\gaps}{5}

\begin{document}
\title{On the Gaussian Interference Channel with Half-Duplex Causal Cognition}

\author{
\IEEEauthorblockN{Martina Cardone$^{\dagger}$, Daniela Tuninetti$^*$, Raymond Knopp$^{\dagger}$, Umer Salim$^{\ddagger}$,}\\
$^{\dagger}$Eurecom,
Biot, 06410, France, 
Email: \{cardone, knopp\}@eurecom.fr\\
$^*$ University of Illinois at Chicago,
Chicago, IL 60607, USA, 
Email: danielat@uic.edu\\
$^{\ddagger}$ Intel Mobile Communications,
Sophia Antipolis, 06560, France, 
Email: umer.salim@intel.com
}
\maketitle
\begin{abstract}
This paper studies the two-user Gaussian interference channel with half-duplex {\em causal cognition}.
This channel model consists of two source-destination pairs sharing a common wireless channel. 
One of the sources, referred to as the {\em cognitive}, 
overhears
the other source, referred to as the {\em primary}, through a noisy link and can therefore assist
in sending the primary's data. Due to practical constraints, the cognitive source is assumed to work in half-duplex mode, that is, it cannot simultaneously transmit and receive.  This model is more relevant for practical cognitive radio systems than the classical information theoretic cognitive channel model, where the cognitive source is assumed to have a non-causal 
knowledge of the primary's message. 
Different network topologies are considered, corresponding to different interference scenarios: (i) the interference-symmetric scenario, where both destinations are in the coverage area of the two sources and 
hence experience interference,
and (ii) the interference-asymmetric scenario, where one destination does not suffer from interference.
For each topology the 
sum-rate performance is studied by first deriving the generalized Degrees of Freedom (gDoF), or ``sum-capacity pre-log'' in the high-SNR regime, and then showing relatively simple coding schemes that achieve a sum-rate upper bound to within a constant number of bits for any SNR.
Finally, the gDoF of the channel is compared to that of the non-cooperative interference channel
and to that of the non-causal cognitive channel
to identify the parameter regimes where half-duplex causal cognition is useless in practice
or  attains its ideal ultimate limit, respectively. 
\end{abstract}


\section{Introduction}
\label{sec:into}
\let\thefootnote\relax\footnote{The results in this paper were presented in part to the 2013 IEEE International Symposium on Information Theory \cite{cardoneCCICISIT2013}.

Manuscript received date: April 1, 2013

Manuscript revised dates: August 20, 2013 and October 30, 2013}
The next major upgrade of fourth generation cellular networks will consist of a massive deployment of radio infrastructure nodes (i.e., base stations and relay stations) enabling different aspects of the cognitive radio paradigm in an operated network scenario \cite{3GPPRel10doc}. Cognitive radio concepts, such as centralized and distributed interference management and collaboration between radio nodes, will allow flexible and multi-band access to the spectrum. Radio infrastructure nodes will come in several flavors, characterized primarily by their available bandwidth and number of concurrent frequency channels on which they can simultaneously operate (spectrum aggregation), the capacity of their backhaul links to the operator's core network (wireless, high throughput/low-latency wired interconnect, non carrier-grade wired backhaul, etc.), their ability to collaborate with other similar nodes, and their coverage area and tolerance to interference. The combined use of  several infrastructure nodes with different features will result in so-called {\em heterogenous networks}. The collaborative features of heterogeneous networks can range from fully-centralized Multiple-Input-Multiple-Output (MIMO) with distributed antennas (very high-quality backhaul links for node connection) achieving full use of the network resources, to distributed (causal) MIMO with improved resource utilization compared to point-to-point channels, or to looser forms of collaboration such as joint time/frequency-allocation for improving link-quality.

In this paper we consider a particular aspect of future heterogeneous networks, namely, a practical application of the cognitive overlay paradigm \cite{goldsmith:spectrymgridlock}. The channel is modeled as a Gaussian two-user Half-Duplex Causal Cognitive Interference Channel (G-HD-CCIC) where one {\em cognitive} source, CTx, 
{overhears}
the other {\em primary} source, PTx, through a noisy channel. The CTx assists 
in sending the PTx's data to the primary 
{destination}, PRx. 
In the language of \cite{3GPPRel10doc}, the PTx could be a macro-base station with a large coverage area and the CTx a small-cell base station or relay station (indoor or a localized coverage-area). The PTx aims to serve a large-number of users, here for simplicity modeled by a single PRx, which may or may not be in the coverage area of CTx. The CTx aims to serve a smaller-number of users, here modeled for simplicity as a single CRx, which may or may not be in the coverage area of PTx. The link between the two sources
is noisy and of limited capacity. We assume that both {sources}
have independently generated messages, each known at the corresponding source only. 
\begin{figure*}
\centering
\includegraphics[width=0.75\textwidth]{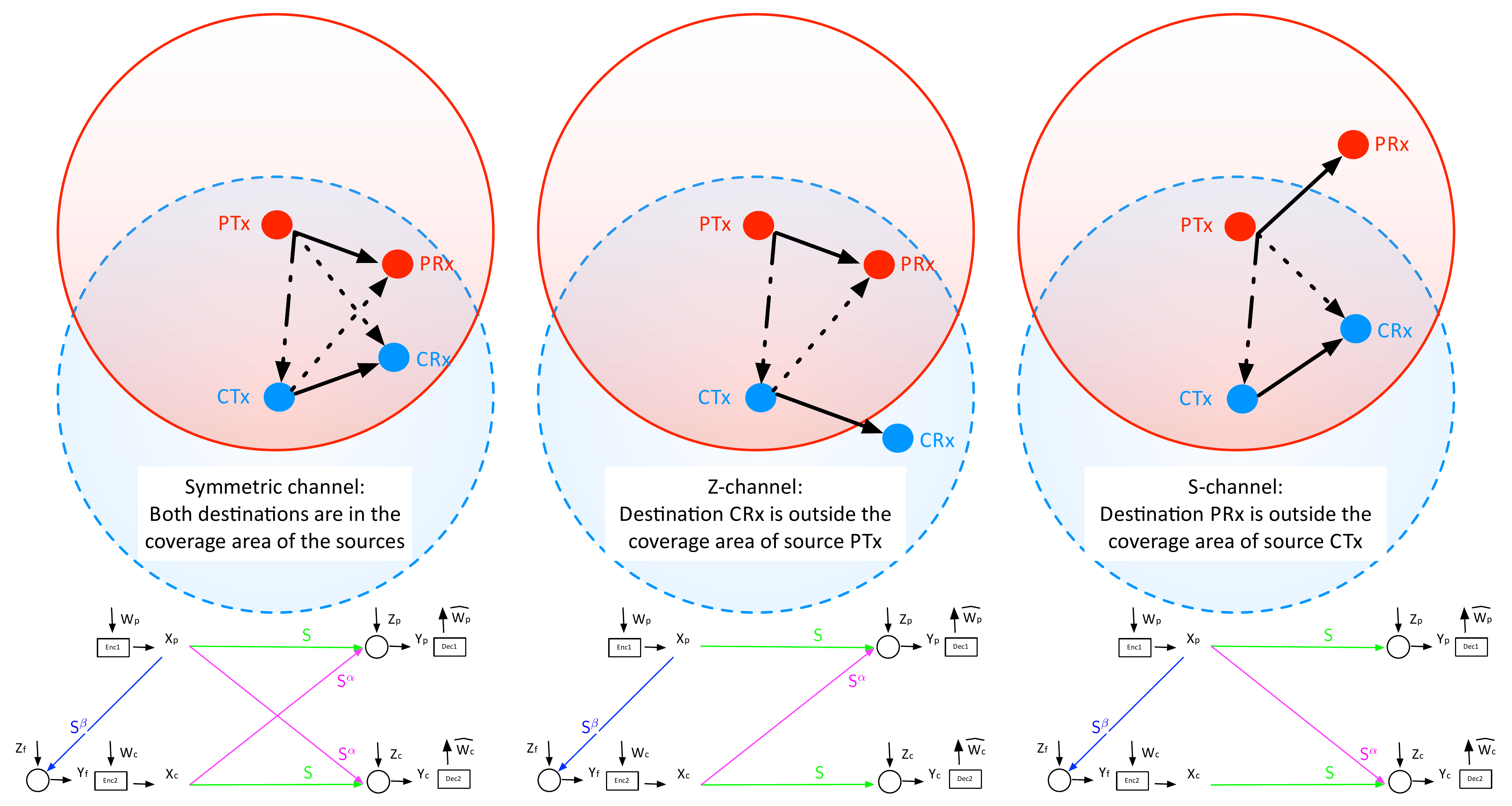}
\caption{The G-HD-CCIC. Left: symmetric channel. Center: $Z$-channel. Right: $S$-channel.}
\label{fig:fig1differenttopologies}
\vspace{-4mm}
\end{figure*}
This corresponds to a scenario where PTx and CTx have separate interconnections to the core network or where the schedulers in PTx and CTx operate in a non-coordinated fashion. We further consider the case where the link from~PTx~to~CTx is unilateral so that CTx can {\em causally} learn the PTx's message but not vice-versa. This models the scenario where the coverage-area of CTx is smaller than that of PTx. 
The final modeling assumption is that CTx operates in {\em half-duplex mode},
meaning that in any time/frequency slot the CTx cannot simultaneously transmit and receive.

We study different deployment configurations, which correspond to different interference scenarios.
In the interference-symmetric scenario (see Fig.~\ref{fig:fig1differenttopologies}, left) both destinations are in the coverage area of the two sources; this implies that both destinations are interfered.
In the interference-asymmetric scenario, one destination does not suffer from interference; in this case one of the interfering links is set to zero. Due to the asymmetry in the cooperation, two interference-asymmetric scenarios must be considered: 
the $Z$-channel, where the link from~PTx~to~CRx is non-existent (i.e., CRx is out of the range of PTx, see Fig.~\ref{fig:fig1differenttopologies}, middle) and
the $S$-channel, where the link from~CTx~to~PRx is non-existent (i.e., PRx is out of the range of CTx, see Fig.~\ref{fig:fig1differenttopologies}, right).
For each topology we study the ultimate sum-rate performance in the spirit of \cite{etw}. We first derive, in closed form, the generalized Degrees-of-Freedom (gDoF), or ``sum-capacity pre-log'' in the high-SNR regime, and we then show relatively simple coding schemes that achieve the sum-rate upper bound to within a constant number of bits regardless of the SNR.

\subsection{Related Work}

{\bf Non-causal CIC:}
In the classical information theoretic cognitive radio overlay paradigm \cite{goldsmith:spectrymgridlock}, the CTx is assumed to have non-causal, i.e., before transmission begins, knowledge of the message and codebook of the PTx \cite{Devroye}. The capacity region of this system in Gaussian noise is known exactly for some channel parameters and to within $1$~bit otherwise \cite{riniJ1}. 

The assumption of a priori knowledge at the CTx is too idealistic for practical schemes and reasonable only in certain situations, such as when the PTx and the CTx are close to each other. Motivated by this practical issue, we assume the existence of a noisy link between the two sources
through which the CTx can causally learn the PTx's message. 
This model is an IC with source cooperation, more specifically, and IC with unilateral cooperation. As such, the CCIC can be studied within the framework of the IC with source cooperation.

{\bf Full-Duplex IC with bilateral source cooperation:}
The cooperative IC was first studied in \cite{HostMadsenIT06}, where Full-Duplex (FD) bilateral source or destination cooperation was analyzed by developing capacity outer and inner bounds for the Gaussian noise channel. 
In \cite{TuninettiITA10} and \cite{PVIT11}, {outer bound regions} were derived for the case of general source cooperation.
In \cite{YANG-TUNINETTI}, an achievable scheme for the IC with FD bilateral source cooperation was proposed by {using rate-splitting \cite{HK} and} Dirty-Paper-Coding (DPC) \cite{costaDPC}.
For FD bilateral source cooperation, \cite{PVIT11} characterized the Gaussian sum-capacity to within 20~bits for all channel parameters {by assuming symmetric, i.e., same strength, cooperation links}, while \cite{YangHighCoop} to within 4~bits in the case of ``strong cooperation'' {with symmetric cooperation, direct and interfering links}. 

{\bf Full-Duplex IC with unilateral source cooperation:}
FD unilateral source cooperation is a special case of FD bilateral source cooperation in which only one of the sources cooperates. This model was studied in \cite{MirmohseniIT2012} for the case of ``look ahead'' at the cognitive source, meaning that at a given time instant CTx has non-causal access to $L\geq 0$ future channel outputs. 
The authors of \cite{MirmohseniIT2012} derived potentially tighter outer bounds for the FD-CCIC channel (i.e., case $L=0$) than those of \cite{TuninettiITA10} specialized to unilateral cooperation; unfortunately it is not clear how to evaluate these bounds in Gaussian noise because they are expressed as a function of auxiliary random variables jointly distributed with the inputs. The achievable region in \cite[Theorem 5 / Corollary 1]{MirmohseniIT2012} is also no smaller than that of \cite{YANG-TUNINETTI} specialized to the case of unilateral cooperation (see \cite[Remark 2, point 6]{MirmohseniIT2012}); unfortunately 
its evaluation in general is very involved as the rate region is specified by 9 jointly distributed auxiliary random variables and by 30 rate constraints.
In \cite{MirmohseniIT2012} inner bounds were evaluated numerically, 
but a performance guarantee in terms of (sum-)capacity to within a constant gap was not given. 

In \cite{cardoneJ1}, the capacity region of the IC with FD source cooperation was characterized to within 2 bits for a large set of channel parameters that, roughly
speaking, excludes the case of weak interference at both receivers.
Although the FD-CCIC represents a more realistic model for practical cognitive radio systems than the non-causal CIC, the FD assumption at the 
CTx has practical restrictions such as the inability to perfectly cancel the self-interference. Due to this practical constraint, in this work we assume that the CTx operates in HD mode.

{\bf  Half-Duplex IC with source cooperation:}
HD cooperation can be studied as a special case of FD cooperation by using the formalism of \cite{kramer-allerton}. This approach is usually not followed in the literature, often making imprecise claims about capacity and Gaussian capacity to within a constant gap. In \cite{kramer-allerton}, it was shown that in the relay channel a larger rate can be achieved by randomly switching between the transmit- and receive-phases at the HD relay. In this way, the randomness that lies into the switch can be harnessed to transmit (at most 1~bit per channel use of) {additional} information to the destination.  We shall refer to this
mode of operation as {\em random switch} \cite{kramer-allerton}, as opposed to {\em deterministic switch} where the transmit- and receive-phases are predetermined and therefore known to all nodes. In~\cite{usICC2013} we showed that random switch is necessary to achieve capacity for a class of deterministic HD relay channels.

In \cite{RuiWu}, the sum-capacity of the Gaussian IC with HD source cooperation and deterministic switch was characterized to within 20~bits and 31~bits for the case of {symmetric (direct, interference {and cooperation} links)} bilateral and {general} unilateral cooperation, respectively. These approximately optimal schemes are inspired by the Linear Deterministic Approximation (LDA) of the Gaussian noise channel.

{\bf Deterministic / noiseless channel models:}
The LDA, first proposed in \cite{avestimher:netflow} in the context of relay networks, captures in a simple deterministic way the interaction between interfering signals {of different strengths}. In the LDA the effect of the noise is neglected and the signal interaction is modeled as bit-wise additions. Thereby, this simplification allows for a complete characterization of the capacity region in many instances where the capacity of the noisy channel counterpart is a long standing open problem. More importantly, the capacity achieving schemes for the LDA model can be `translated' into schemes for the Gaussian noise case that, although not optimal in general, are at most a constant and finite number of bits away from an outer bound {regardless of the channel parameters}.
For example, in the non-cooperative IC, the evaluation of the Han-Kobayashi region \cite{HK} in the Gaussian noise case is a formidable task \cite{etw}, which does not appear to provide useful insights in its full generality; the LDA model instead reveals the (approximate) optimality of the simple rate-splitting method proposed in \cite{etw}, namely that, whatever is treated as noise should be received at the level of the noise \cite{etw}; this strategy does not exactly achieve capacity, but it is 
optimal to within $1$~bit.
The ``it is better to approximate''-philosophy \cite{TsePlenaryISIT2009} behind the LDA model has enabled great progress in long-standing open problems, including the non-cooperative IC \cite{etw} and the relay channel \cite{avestimher:netflow}. 

\subsection {Contributions} 
This work characterizes both the interference-symmetric and interference-asymmetric sum-capacity to within a constant gap, hence the gDoF, of the G-HD-CCIC. 
Our main contributions can be summarized as follows:

$\bullet$ We overcome a number of limitations and improve on the results of \cite{RuiWu} as follows: 
(i) we properly account for random switch at the CTx in the outer bound by using the framework of \cite{kramer-allerton}, i.e., without the need to develop a separate theory for the HD case as in \cite{RuiWu},
(ii) we consider the classical definition of sum-capacity without introducing any ``back-off'' in the PTx rate, which can be interpreted as a sort of interference margin at the PRx,
(iii) we derive the gDoF in closed form rather than expressing it implicitly as the solution of a linear program, and
(iv) we reduce the gap to \gap~bits for the interference-symmetric case, and to \gapz~bits and \gaps~bits for the $Z$-~and $S$-channels, respectively, through novel achievable schemes. 

$\bullet$ As in \cite{RuiWu}, our `optimal to within a constant gap' schemes for the G-HD-CCIC are inspired by the LDA model. Using the LDA model, we obtain a closed-form expression for the different optimization variables (e.g., schedule, power splits, coding schemes and corresponding decoding orders, etc.). This result sheds light on how the design of the HD CTx should be properly carried out, which is an important practical task for future wireless networks. 

$\bullet$ We compare the gDoF of the G-HD-CCIC with that of:
(i) the classical non-cooperative IC, i.e., where there is no cooperation among the nodes \cite{etw}, 
and (ii) the non-causal cognitive IC, i.e., where the CTx has a non-causal knowledge of the PTx's message \cite{riniJ1}. In particular, we find the parameter regimes where HD unilateral cooperation does not yield benefits compared to the non-cooperative IC \cite{etw}, and those where it attains the ultimate performance limits of the non-causal CIC \cite{riniJ1}. Interestingly, we show that in regimes where the G-HD-CCIC outperforms the non-cooperative IC the cooperation link must be able to reliably convey a rate larger than the sum-capacity of the corresponding non-cooperative IC. 
{Moreover, we identify the regimes where a loss incurs by using HD mode of operation at the CTx with respect to FD \cite{cardoneJ1}.}

\subsection{Paper Organization}
The rest of the paper is organized as follows. 
Section~\ref{sec:model} describes the channel model and defines the sum-capacity to within a constant gap and the gDoF.
Section~\ref{sec:outbound} derives an outer bound for our HD model by adapting known FD bounds.
Sections~\ref{sec:symm},~\ref{sec:asymm Z} and~\ref{sec:asymm S} characterize the sum-capacity to within a constant gap, and hence the gDoF, for the interference-symmetric, the $Z$- and $S$-channels, respectively. 
Section~\ref{sec:Concl} concludes the paper.
The details of the proofs can be found in Appendix.

\section{Channel Model}
\label{sec:model}
We use the notation convention of \cite{book:ElGamalKim2012}: $[n_1:n_2]$ is the set of integers from $n_1$ to $n_2 \geq  n_1$, $[x]^+ := \max\{0,x\}$ for $x\in\mathbb{R}$, and $Y^{j}$ is a vector of length $j$ with component $(Y_1,\ldots,Y_j)$.

We consider the single-antenna G-HD-CCIC, whose input/output relationship is given by \eqref{eq:chmodel} at the top of next page.
\begin{figure*}
\begin{align}
\begin{bmatrix}
\mathsf{Y}_\mathsf{f} \\ \mathsf{Y}_\mathsf{p} \\ \mathsf{Y}_\mathsf{c} \\ 
\end{bmatrix}
=
\begin{bmatrix}
1-\mathsf{M}_{\mathsf{c}} & 0 & 0 \\
0 & 1 & 0 \\
0 & 0 & 1 \\
\end{bmatrix}
\begin{bmatrix}
\sqrt{\mathsf{C}} & \star  \\
\sqrt{\mathsf{S}_\mathsf{p}} & \sqrt{\mathsf{I}_\mathsf{c}} \ {\rm e}^{{\rm j}\theta_{\mathsf{c}}} \\
\sqrt{\mathsf{I}_\mathsf{p}} \ {\rm e}^{{\rm j}\theta_{\mathsf{p}}} & \sqrt{\mathsf{S}_\mathsf{c}} \\
\end{bmatrix}
\begin{bmatrix}
1 & 0 \\
0 & \mathsf{M}_{\mathsf{c}} \\
\end{bmatrix}
\begin{bmatrix}
\mathsf{X}_\mathsf{p} \\ \mathsf{X}_\mathsf{c} \\
\end{bmatrix}
+
\begin{bmatrix}
\mathsf{Z}_\mathsf{f} \\ \mathsf{Z}_\mathsf{p} \\ \mathsf{Z}_\mathsf{c} \\ 
\end{bmatrix}
\label{eq:chmodel}
\end{align}
\vspace{-4mm}
\end{figure*}
The channel inputs are subject, without loss of generality, to the average power constraint
$\mathbb{E} \left [ |\mathsf{X}_{i}|^2 \right ]  \leq  1$, $i   \in   \{\mathsf{p},\mathsf{c}\}$
(i.e., non-unitary power constraints can be incorporated into the channel gains).
$\mathsf{M}_{\mathsf{c}}$ is the state random variable that indicates whether the CTx is in receive-mode ($\mathsf{M}_{\mathsf{c}}=0$) or in transmit-mode ($\mathsf{M}_{\mathsf{c}}=1$) \cite{kramer-allerton}.
A $\star$  in the channel transfer matrix 
indicates the channel gain that does not affect the capacity region because of the HD constraint.
The channel parameters $(\mathsf{C},\mathsf{S}_\mathsf{p},\mathsf{S}_\mathsf{c},\mathsf{I}_\mathsf{p},\mathsf{I}_\mathsf{c},\theta_{\mathsf{p}},\theta_{\mathsf{c}})\in\mathbb{R}_+^{7}$ are fixed and so known to all nodes.
Some of the channel gains can be taken to be real-valued and non-negative since a node can compensate for the phase of one of its channel gains. In the following we assume that the channel sub-matrix 
${\tiny \begin{bmatrix}
\sqrt{\mathsf{S}_\mathsf{p}} & \sqrt{\mathsf{I}_\mathsf{c}} \ {\rm e}^{{\rm j}\theta_{\mathsf{c}}} \\
\sqrt{\mathsf{I}_\mathsf{p}} \ {\rm e}^{{\rm j}\theta_{\mathsf{p}}} & \sqrt{\mathsf{S}_\mathsf{c}} \\
\end{bmatrix}}$
is full-rank (otherwise one channel output is a degraded version of the other and hence one receiver can, without loss of generality, decode all messages).
The noises are independent proper-complex Gaussian random variables with, without loss of generality, zero mean and unit variance.

PTx has a message $\mathsf{W}_\mathsf{p} \in  [1:2^{N {R}_\mathsf{p}}]$ for PRx and CTx has a message $\mathsf{W}_\mathsf{c}\in [1:2^{N {R}_\mathsf{c}}]$ for CRx, where $N$ denotes the codeword length and ${R}_\mathsf{p}$ and ${R}_\mathsf{c}$ are the transmission rates for PTx and CTx, respectively, measured in bits per channel use.
The messages $\mathsf{W}_\mathsf{p}$ and $\mathsf{W}_\mathsf{c}$ are independent and uniformly distributed on their respective domains. At time $t$, $t \in [1:N]$, PTx maps its message $\mathsf{W}_\mathsf{p}$ into a channel input symbol according to $\mathsf{X}_{\mathsf{p},t}\left ( \mathsf{W}_\mathsf{p} \right )$ and CTx encodes its message $\mathsf{W}_\mathsf{c}$ and its past channel observations into $\mathsf{X}_{\mathsf{c},t}\left ( \mathsf{W}_\mathsf{c},  \mathsf{Y}_{\mathsf{f}}^{t-1}\right )$. 
The channel is assumed to be memoryless.
At time $N$, PRx outputs an estimate of $\mathsf{W}_\mathsf{p}$ based on all its channel observations as
$\widehat{\mathsf{W}}_\mathsf{p}\left ( \mathsf{Y}_\mathsf{p}^N \right )$ and similarly CRx outputs $\widehat{\mathsf{W}}_\mathsf{c}\left ( \mathsf{Y}_\mathsf{c}^N \right )$.
The capacity region is the convex closure of all non-negative rate pairs $\left ( {R}_\mathsf{p}, {R}_\mathsf{c} \right )$ such that
$\max_{i\in\{\mathsf{p},\mathsf{c}\}} 
\mathbb{P} \Big[ \widehat{\mathsf{W}}_i \neq \mathsf{W}_i \Big]
\rightarrow 0$ as $N \rightarrow +\infty$.
  
The exact capacity of the CCIC is open.
We make progress towards understanding the ultimate performance limits of the G-HD-CCIC by {\em approximately} characterizing its sum-capacity in the spirit of \cite{etw}. The sum-capacity is known to within $\mathsf{GAP}$ bits if one can show a sum-rate inner bound $({R}_\mathsf{p}+{R}_\mathsf{c})^{\rm(IB)}$ and a sum-rate outer bound $({R}_\mathsf{p}+{R}_\mathsf{c})^{\rm(OB)}$ such that 
$
 ({R}_\mathsf{p}+{R}_\mathsf{c})^{\rm(OB)} - ({R}_\mathsf{p}+{R}_\mathsf{c})^{\rm(IB)} \leq \mathsf{GAP},
$
where $\mathsf{GAP}$ is a constant with respect to the channel parameters.

The knowledge of the sum-capacity to within a constant gap implies the {\em exact} knowledge of the {\em sum-capacity pre-log factor at high SNR} \cite{etw}. By following \cite[Section V]{etw}, for some $\mathsf{SNR}>0$ and for some non-negative $(\alpha_\mathsf{c},\alpha_\mathsf{p},\beta)$, we let
\begin{align}
  &\mathsf{S}_\mathsf{c} = \mathsf{S}_\mathsf{p} {= \mathsf{S}} := \mathsf{SNR}^{1},
\ \mathsf{C}             := \mathsf{SNR}^{\beta} \nonumber
\\& \mathsf{I}_\mathsf{c}  := \mathsf{SNR}^{\alpha_\mathsf{c}}, \ 
\ \mathsf{I}_\mathsf{p}  := \mathsf{SNR}^{\alpha_\mathsf{p}}, \ 
\label{eq:par}
\end{align}
where $\alpha_i$ is the ratio of the received power on the interference link~$i\in\{\mathsf{p},\mathsf{c}\}$ expressed  in dB  over the received power on the direct link expressed in dB, {and where $\beta$ is defined similarly for the cooperation link. Note that the direct links are assumed here to have the same strength in order to reduce the number of parameters.
The gDoF, following \cite{etw}, is defined as 
\begin{align}
\mathsf{d} &:= 
\lim_{\mathsf{SNR}\to+\infty} \frac{\max\{{R}_\mathsf{p} + {R}_\mathsf{c}\}}{2\log(1+ \mathsf{SNR})}
\label{eq:gDoF}
\end{align}
where the maximization is intended over all possible achievable rates $({R}_\mathsf{p},{R}_\mathsf{c})$. 
The normalization in~\eqref{eq:gDoF} of the maximum sum-rate is with respect to the sum-capacity of an interference-free network.
In the non-cooperative case $\mathsf{d}^{\rm(NoCoop)} \leq 1$
because the absence of interference is the best possible scenario.
In the cooperative case, interference can be beneficial because it can carry useful cooperative information \cite{PVIT11,YangHighCoop} and, as a result, $\mathsf{d}$ in~\eqref{eq:gDoF} can be larger than 1. In the limiting ideal case, where the CTx non-causally knows the PTx's message, the gDoF, indicated as $\mathsf{d}^{\rm(Ideal)}$, can grow with the interference exponent $\alpha$.
In general, $\mathsf{d}^{\rm(NoCoop)} \leq \mathsf{d} \leq \mathsf{d}^{\rm(Ideal)}$.

In this work we derive results for both the interference-symmetric and interference-asymmetric cases. In particular,
the interference-symmetric channel has $\alpha_\mathsf{p}=\alpha_\mathsf{c}=\alpha$,
the $Z$-channel has $\alpha_\mathsf{c}=\alpha$ and $\alpha_\mathsf{p}=0$, and
the $S$-channel has $\alpha_\mathsf{p}=\alpha$ and $\alpha_\mathsf{c}=0$, for some $\alpha\geq 0$. 
For these cases, the gDoF in~\eqref{eq:gDoF} is a function of $(\alpha,\beta)$; we shall refer to $\alpha$ as the `interference exponent' and to $\beta$ as the `cooperation exponent'. We expect that $\mathsf{d}^{\rm(NoCoop)} = \mathsf{d}(\alpha,0)$ and 
 $\mathsf{d}^{\rm(Ideal)} = \lim_{\beta\to\infty} \mathsf{d}(\alpha,\beta)$.

\section{Sum-capacity Upper Bound}
\label{sec:outbound}
Here we specialize the known outer bounds for FD bilateral source cooperation in \cite{HostMadsenIT06,TuninettiITA10,PVIT11} to the case of HD unilateral cooperation by following the approach of~\cite{kramer-allerton}. We can show that \eqref{eq:upupup}, at the top of next page, holds by proceeding through the following steps:
\begin{figure*}
\begin{align}
\left({R}_\mathsf{p}+{R}_\mathsf{c} \right)^{\rm(OB)}
&:= \min\Big\{
\left({R}_\mathsf{p}+{R}_\mathsf{c} \right)^{\rm(CS)},
\left({R}_\mathsf{p}+{R}_\mathsf{c} \right)^{\rm(DT)},
\left({R}_\mathsf{p}+{R}_\mathsf{c} \right)^{\rm(PV)} \Big\}
\quad \text{for}
\label{eq:upupup}
\\
\left( {R}_\mathsf{p}+{R}_\mathsf{c} \right)^{\rm(CS)} 
&:= 
2.507+ \min \Big\{ \gamma \log \left( 1+\mathsf{S}+\mathsf{C} \right)+2\left( 1-\gamma \right)\log \left( 1+\mathsf{S}\right), 
\gamma \log \left( 1 + \mathsf{S} + \mathsf{I}_\mathsf{p} \right) +
\nonumber \\&
+ \left( 1 - \gamma\right)\log \left( 1 + \left( \sqrt{\mathsf{S}} + \sqrt{\mathsf{I}_\mathsf{c}}\right )^2 + \left( \sqrt{\mathsf{S}} + \sqrt{\mathsf{I}_\mathsf{p}}\right )^2 + \left | \mathsf{S} +\sqrt{\mathsf{I}_\mathsf{p}\mathsf{I}_\mathsf{c}} \right |^2 \right)\Big\}
\label{eq:CutSet} 
\\
\left( {R}_\mathsf{p} + {R}_\mathsf{c} \right)^{\rm(DT)}  
&:=   \min  \left \{ 3 +  \gamma \log  \left( 1  +  \mathsf{S} \right) +  
  \left( 1 - \gamma \right) \log  \left( \frac{\max \left \{ \mathsf{I}_\mathsf{c},\mathsf{S} \right \}}{\mathsf{I}_\mathsf{c}} \cdot
\left(  1   +  \left( \sqrt{\mathsf{S}}  +  \sqrt{\mathsf{I}_\mathsf{c}}\right)^2 \right) \right) ,\right. 
\nonumber 
\\& \left. 2 +  \gamma \log   \left( 1 + \mathsf{C} + \max \left\{\mathsf{I}_\mathsf{p},\mathsf{S}\right \} \right) 
+ \left( 1 - \gamma \right) \log  \left(\frac{\max \left \{ \mathsf{I}_\mathsf{p},\mathsf{S} \right \}}{\mathsf{I}_\mathsf{p}} \cdot
\left( 1 + \left( \sqrt{\mathsf{S}} + \sqrt{\mathsf{I}_\mathsf{p}}\right)^2 \right)  \right)  \right \}
\label{eq:Tuninetti}
\\ 
\left( {R}_\mathsf{p}+{R}_\mathsf{c} \right)^{\rm(PV)}
&:= {3.5048}+\gamma \log \left( 1+\mathsf{S}+\mathsf{C}+\mathsf{I}_\mathsf{p} \right)+\left( 1-\gamma \right)\log \left( 1+\mathsf{I}_\mathsf{p}+\frac{\mathsf{S}}{\mathsf{I}_\mathsf{c}} \right) + 
\nonumber \\&
+\left( 1-\gamma \right)\log \left( 1+\mathsf{I}_\mathsf{c}+\frac{\mathsf{S}}{\mathsf{I}_\mathsf{p}} \right).
\label{eq:Prabh}
\end{align}
\vspace{-7mm}
\end{figure*}
(i) in the outer bounds for the general memoryless IC with bilateral FD source cooperation, 
we substitute $\mathsf{X}_\mathsf{c}$ with the pair $(\mathsf{X}_\mathsf{c},\mathsf{M}_{\mathsf{c}})$~\cite{kramer-allerton};
(ii) for any triplet of random variables $(\mathsf{A},\mathsf{B},\mathsf{C})$ we bound  $I(\mathsf{A},\mathsf{X}_\mathsf{c},\mathsf{M}_{\mathsf{c}};\mathsf{B}|\mathsf{C}) \leq H(\mathsf{M}_{\mathsf{c}})+I(\mathsf{A},\mathsf{X}_\mathsf{c}; \mathsf{B}|\mathsf{C},\mathsf{M}_{\mathsf{c}})$ since, for a binary-valued random variable $\mathsf{M}_{\mathsf{c}}$, we have $I(\mathsf{M}_{\mathsf{c}};\mathsf{B}|\mathsf{C}) \leq H(\mathsf{M}_{\mathsf{c}})  = \mathcal{H}(\gamma)$ for some $\gamma:=\mathbb{P}[\mathsf{M}_{\mathsf{c}}=0]\in[0,1]$ 
and where  
$\mathcal{H}(\gamma)= -\gamma\log(\gamma)-(1-\gamma)\log(1-\gamma)$;
(iii) for all the remaining mutual information terms, which are conditioned on $\mathsf{M}_{\mathsf{c}}=\ell$, $\ell\in[0:1]$, the ``Gaussian maximizes entropy''-principle guarantees that in order to exhaust all possible input distributions it suffices to consider jointly Gaussian  proper-complex inputs with covariance matrix
${\tiny \begin{bmatrix}
P_{\mathsf{p},\ell} & \rho_{\ell} \ \sqrt{P_{\mathsf{p},\ell}P_{\mathsf{c},\ell}} \\ 
\rho_{\ell}^* \ \sqrt{P_{\mathsf{p},\ell}P_{\mathsf{c},\ell}} & P_{\mathsf{c},\ell}\\
\end{bmatrix}}$
for $|\rho_\ell|\leq 1$ and $(P_{\mathsf{p},0},P_{\mathsf{p},1},P_{\mathsf{c},0},P_{\mathsf{c},1}) \in \mathbb{R}^4_{+}$ satisfying the power constraint $\gamma P_{u,0} +  \left( 1-\gamma\right)P_{u,1} \leq  1, \ u \in \left \{ \mathsf{p},\mathsf{c}\right \}$;
(iv) since PTx always transmits we define, for some $\tau \in \left [ 0,1 \right ]$, $P_{\mathsf{p},0}=\frac{\tau}{\gamma}$ and $P_{\mathsf{p},1}=\frac{1-\tau}{1-\gamma}$, while, since the CTx transmission only affects the receivers outputs when $\mathsf{M}_{\mathsf{c}}=1$, we let $P_{\mathsf{c},0}=0$ and $P_{\mathsf{c},1}=\frac{1}{1-\gamma}$;
(v) the sum-rate upper bound
in~\eqref{eq:CutSet} is from~\cite[$\min \{\text{eq.(81)+eq.(82)},\text{eq.(83)} \}$]{HostMadsenIT06}, 
that in~\eqref{eq:Tuninetti} is from~\cite[$\min \{ \text{eq.(4d)},\text{eq.(4e)}\}$]{TuninettiITA10}, and 
that in~\eqref{eq:Prabh} is from~\cite[eq.(6)-(7)]{PVIT11} and are obtained by upper bounding each mutual information term over $(\rho_0,\rho_1,\tau) \in  [0,1]^3$, {as well as over the phases of the channel gains,} while keeping $\gamma$ as fixed;
(vi) each outer bound is also characterized by a linear combination of terms of the type `$\gamma \log \left( \gamma \right)$', which can be maximized over $\gamma \in [0,1]$ thus giving the additive constants in~\eqref{eq:CutSet},~\eqref{eq:Tuninetti} and~\eqref{eq:Prabh}.

\section{The Interference-Symmetric G-HD-CCIC}
\label{sec:symm}
When the interfering links have the same strength $\mathsf{I}_\mathsf{c}=\mathsf{I}_\mathsf{p}=\mathsf{SNR}^{\alpha}$, 
we can show
\begin{theorem}
\label{thm:dofuniHDIFC}
The sum-capacity upper bound in \eqref{eq:upupup} is achievable to within \gap \ bits  regardless of the actual value of the channel parameters for the interference-symmetric G-HD-CCIC.
Therefore, the gDoF can be obtained from~\eqref{eq:upupup} and equals
\begin{subequations}
\begin{align}
&\mathsf{d}^{\rm(SYM)}
=\max_{\gamma\in[0,1]} \frac{1}{2}\min 
   \Big\{ \gamma \max \left \{1,\beta \right \}+2\left( 1-\gamma \right), 
\label{eq:gdof sym cs}
\\& \gamma +\left( 1- \gamma \right ) \left( \max \left \{ 1,\alpha \right\}+ \left [ 1-\alpha \right ]^+ \right ), 
\label{eq:gdof sym dt}
\\& \gamma \max \left \{\alpha,\beta,1 \right \}+2\left( 1- \gamma \right ) \max \left \{\alpha,1-\alpha \right \}
\Big\}
\label{eq:gdof sym pv}
\\&=\left\{\begin{array}{ll}
1-\alpha+\frac{1}{2}\frac{[\beta-2+2\alpha]^+ \ \alpha}{\beta+\alpha-1} & \alpha\in[0,1/2) \\
  \alpha+\frac{1}{2}\frac{[\beta-2\alpha]^+ \ (2-3\alpha)}{\beta-3\alpha+1} & \alpha\in[1/2,2/3) \\
\max\left\{1-\frac{1}{2}\alpha,\frac{1}{2}\alpha \right\} & \alpha\in[2/3,2) \\
1+\frac{1}{2}\frac{[\beta-2]^+ \ (\alpha-2)}{\beta + \alpha-3} & \alpha\in[2,\infty) \\
\end{array}.\right.
\label{eq:gdof sym final}
\end{align}
\end{subequations}
\end{theorem}
\begin{IEEEproof}
The details of the proof can be found in Appendix~\ref{sec:constGap}.
\end{IEEEproof}

The gDoF expression in~\eqref{eq:gdof sym final}, to be compared with
$\mathsf{d}^{\rm(NoCoop)} = \min\{1, \max\{1 - \alpha,\alpha\}, \max\{1-\alpha/2,\alpha/2\}\}$,
and
$\mathsf{d}^{\rm(Ideal)}  = \max\{1-\alpha/2,\alpha/2\}$,
has an interesting interpretation, which we shall discuss in details for the different interference regimes in the following.
Before we do so, we introduce the LDA model, or deterministic high-SNR approximation of a Gaussian noise channel  \cite{avestimher:netflow}, which will help us to develop insights into approximately optimal achievable schemes.
The {\em symmetric} LDA has input/output relationship
\begin{align*}
&\mathsf{Y}_\mathsf{f} = (1-\mathsf{M}_{\mathsf{c}}) \ \mathbf{S}^{n-n_\mathsf{f}} \mathsf{X}_\mathsf{p}
\\&
\mathsf{Y}_\mathsf{p} = \mathbf{S}^{n-n_\mathsf{d}} \mathsf{X}_\mathsf{p} + \mathsf{M}_{\mathsf{c}} \ \mathbf{S}^{n-n_\mathsf{i}} \mathsf{X}_\mathsf{c}
\\&
\mathsf{Y}_\mathsf{c} = \mathbf{S}^{n-n_\mathsf{i}} \mathsf{X}_\mathsf{p} + \mathsf{M}_{\mathsf{c}} \ \mathbf{S}^{n-n_\mathsf{d}} \mathsf{X}_\mathsf{c}
\end{align*}
where: $(n_\mathsf{d},n_\mathsf{f},n_\mathsf{i})$ are non-negative integers with $n := \max\{n_\mathsf{f},n_\mathsf{d},n_\mathsf{i}\}$, 
$\mathsf{M}_{\mathsf{c}}$ is the binary random variable that indicates the state of CTx, all input and output vectors have length $n$ and take value in GF(2), the sum is understood bit-wise on GF(2), and $\mathbf{S}$ is the down-shift matrix of dimension $n$. The model has the following interpretation. The PTx sends a length-$n$ vector $\mathsf{X}_\mathsf{p}$, whose top $n_\mathsf{d}$ bits are received at the PRx through the direct link, the top $n_\mathsf{i}$ bits are received at the CRx through the interference link, and the top $n_\mathsf{f}$ bits are received at the CTx through the cooperation/feedback link; similarly for $\mathsf{X}_\mathsf{c}$.
The fact that only a certain number of bits are observed at a given node is a consequence of the `down shift' operation through the matrix $\mathbf{S}$. The  bits not observed at a node are said to be `below the noise floor'.
The parameters of the LDA model can be related to those of the G-HD-CCIC in~\eqref{eq:par} as 
\begin{align*}
&n_\mathsf{d} = \lfloor\log(1+\mathsf{S}_\mathsf{c})\rfloor = \lfloor\log(1+\mathsf{S}_\mathsf{p})\rfloor
\\&
n_\mathsf{i} = \lfloor\log(1+\mathsf{I}_\mathsf{c})\rfloor = \lfloor\log(1+\mathsf{I}_\mathsf{p})\rfloor
\\&
n_\mathsf{f} = \lfloor\log(1+\mathsf{C})\rfloor.
\end{align*}
In the symmetric case we indicate $\alpha := n_\mathsf{i}/n_\mathsf{d}$ and $\beta := n_\mathsf{f}/n_\mathsf{d}$ as they play the same role of the corresponding quantities in~\eqref{eq:par}. The simplicity of the LDA model allows for the exact capacity characterization in many instances where the capacity of the Gaussian counterpart is open. Moreover, the sum-capacity of the LDA normalized by $2 n_\mathsf{d}$ equals the Gaussian gDoF defined in~\eqref{eq:gDoF} in \cite{RuiWu,PVIT11,YangHighCoop}.
The difference between our model and that of \cite{avestimher:netflow} is the inclusion of the binary-valued random variable $\mathsf{M}_{\mathsf{c}}$ to indicate the state of CTx~\cite{kramer-allerton}.
}

\begin{figure*}
\centering
\subfigure[Phase~1 ($\mathsf{M}_{\mathsf{c}}=0$) common to the the symmetric and asymmetric channels.]{
\includegraphics[width=0.62\columnwidth]{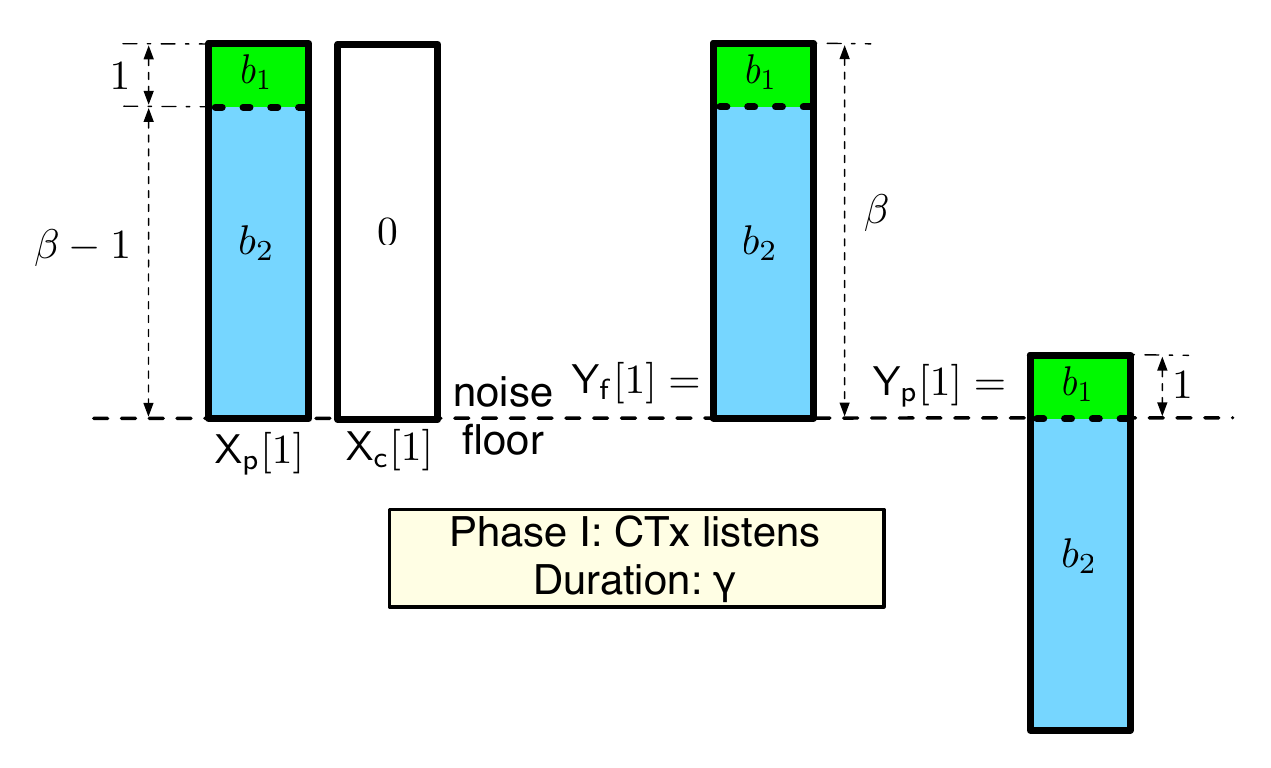}
\label{fig:firstphase}
}
\hfill
\subfigure[Phase~2 ($\mathsf{M}_{\mathsf{c}}=1$) for $\alpha \in [0,1/2)$ for the symmetric channel.]{
\includegraphics[width=0.62\columnwidth]{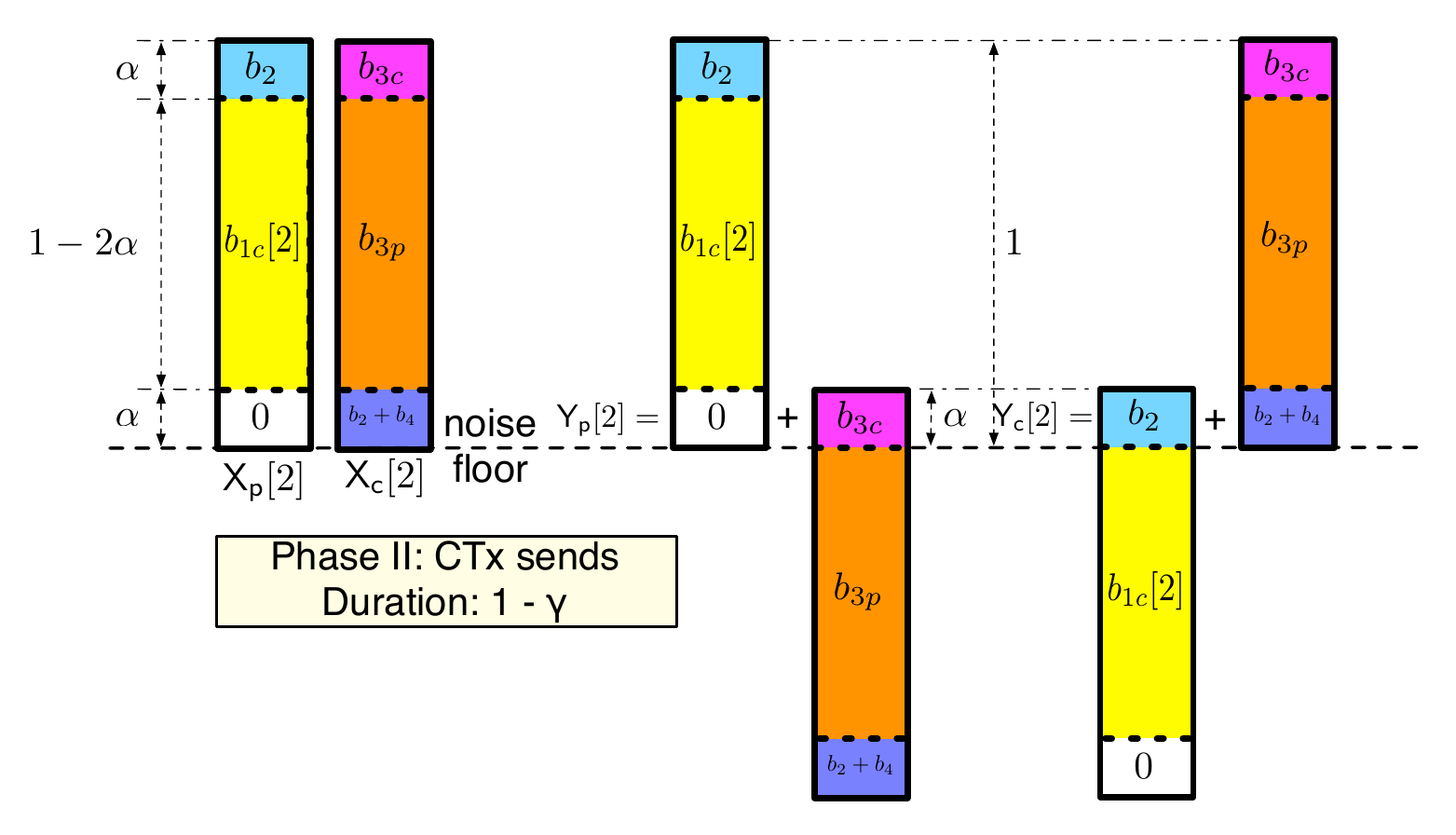}%
\label{fig:secondphase10}
}
\hfill
\subfigure[Phase~2 ($\mathsf{M}_{\mathsf{c}}=1$) for $\alpha \in [1/2,2/3)$ for the symmetric channel.]{
\includegraphics[width=0.62\columnwidth]{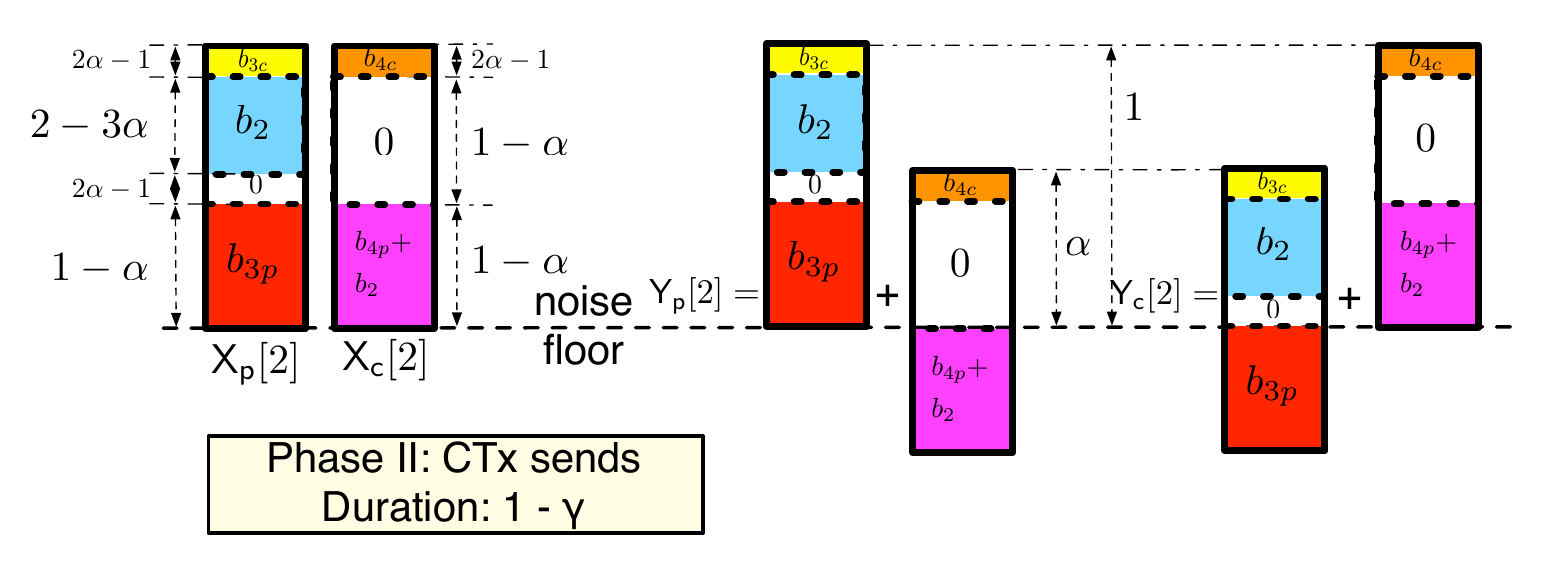}
\label{fig:secondphase8}
}
\hfill
\subfigure[Phase~2 ($\mathsf{M}_{\mathsf{c}}=1$) for $\alpha \in [2,\infty)$ for the symmetric channel.]{
\includegraphics[width=0.62\columnwidth]{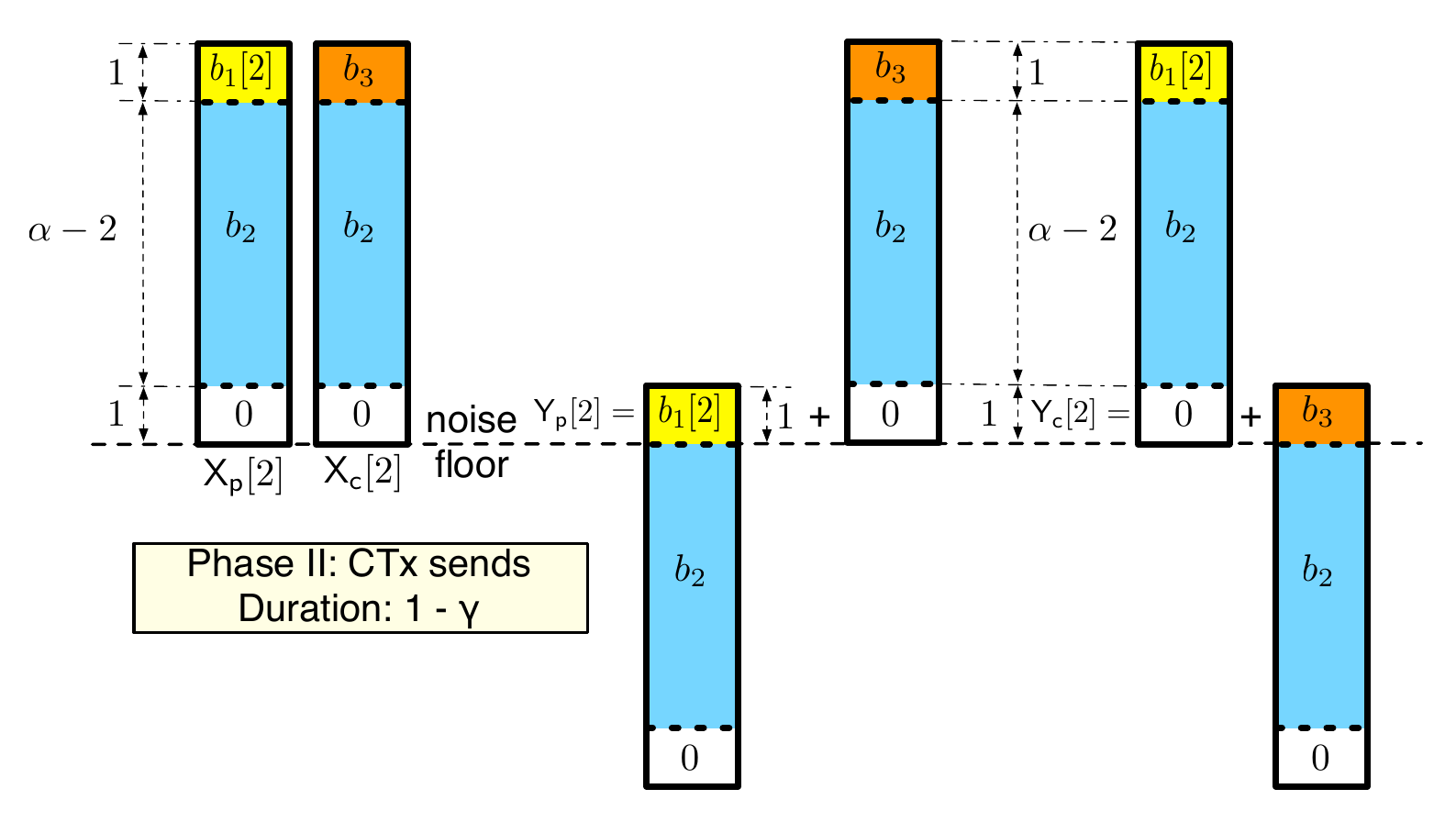}
\label{fig:secondphase3}
}
\hfill
\subfigure[Phase~2 ($\mathsf{M}_{\mathsf{c}}=1$)  for $\alpha \in [0,1)$ for the S-channel.]{
\includegraphics[width=0.62\columnwidth]{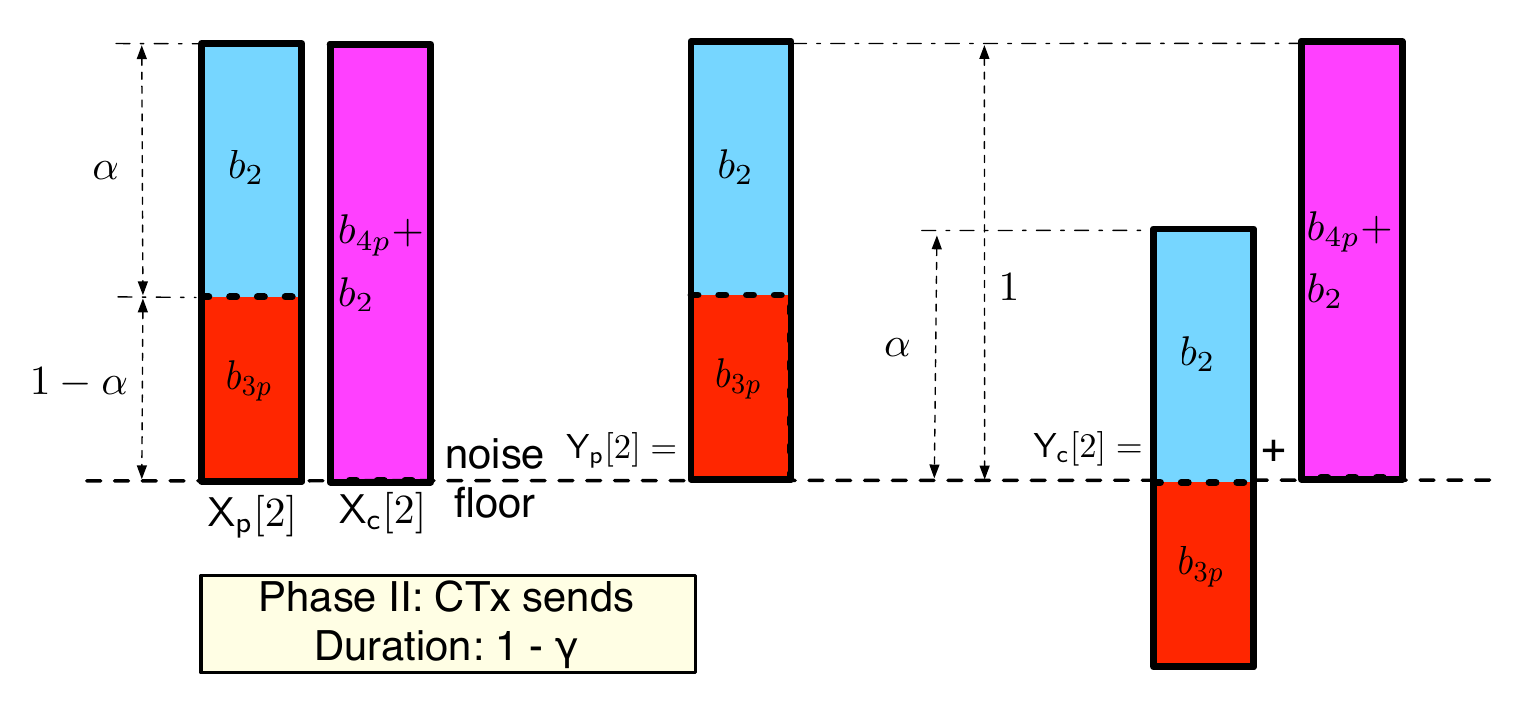}
\label{fig:secondphase5S}
}
\hfill
\subfigure[Phase~2 ($\mathsf{M}_{\mathsf{c}}=1$)  for $\alpha \in [1,2)$ for the S-channel.]{
\includegraphics[width=0.62\columnwidth]{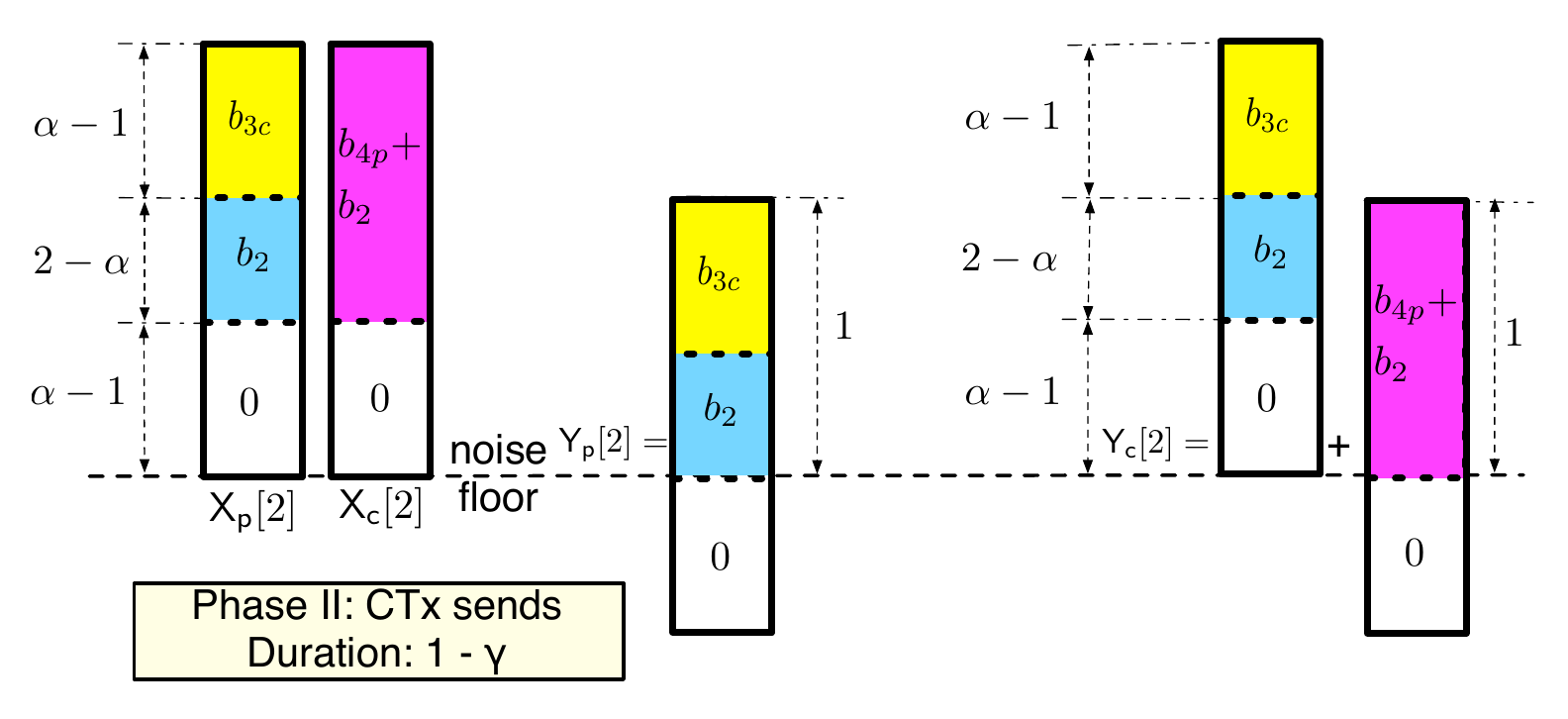}
\label{fig:secondphase4S}
}
\caption{Achievable strategies for the symmetric and asymmetric LDA channels.}
\label{fig:fig5}
\vspace{-4mm}
\end{figure*}

{\bf Very weak interference regime: $\alpha \in [0,2/3)$.}
Without cooperation, i.e., $\beta=0$, the tightest upper bound in this regime is~\eqref{eq:gdof sym pv}~\cite{etw}.
Recall that no-cooperation in equivalent to $\gamma = 0$, {i.e., the cognitive source never listens to the channel.}
For a $\beta > 0$, the bound in~\eqref{eq:gdof sym pv} is optimized by $\gamma = 0$, {that is, by no-cooperation,} whenever
$\max\{1,\alpha,\beta\} \leq 2\max\{1-\alpha,\alpha\} =2\mathsf{d}^{\rm(NoCoop)}$, which is equivalent to $\beta \leq 2\mathsf{d}^{\rm(NoCoop)}$. Intuitions suggest that the cooperation link gain should be ``sufficiently strong'' for HD unilateral cooperation to be beneficial. We can precisely quantify the statement ``sufficiently strong'' as follows:
$\mathsf{d}>\mathsf{d}^{\rm(NoCoop)}$ if $\beta > 2\mathsf{d}^{\rm(NoCoop)}$. Recall that a strict inequality in the gDoF, or sum-capacity pre-log at high SNR, implies that the difference between the sum-capacities with HD unilateral cooperation and without cooperation becomes unbounded when SNR increases. In other words, when the cooperation link can reliably convey a rate larger than the sum-capacity of the non-cooperative IC ($\beta > 2\mathsf{d}^{\rm(NoCoop)}$), HD unilateral cooperation provides an unbounded sum-rate gain compared to the non-cooperative IC.

The optimal $\gamma$ 
is obtained by equating the bounds in~\eqref{eq:gdof sym pv} and~\eqref{eq:gdof sym dt} and is given by
\begin{align}
\gamma^\star = \frac{\min\{2-3\alpha,\alpha\}}{\min\{2-3\alpha,\alpha\}+\beta-1}.
\label{eq:gamma opt sum very weak}
\end{align}

We now give an intuitive argument for the optimal $\gamma$ in~\eqref{eq:gamma opt sum very weak}.
For the case $2-3\alpha\leq\alpha$,  
i.e., $\alpha \in [1/2,2/3)$, an achievable scheme for the LDA is represented in Fig.~\ref{fig:firstphase} and Fig.~\ref{fig:secondphase8} {for the case $\beta \geq 1$}.\footnote{In each sub-figure of Fig.~\ref{fig:fig5}, on the {left} hand side we represent the transmitted signals $\mathsf{X}_\mathsf{p}$ and $\mathsf{X}_\mathsf{c}$, which are vectors of normalized length $n/n_\mathsf{d} = \max\{1,\alpha,\beta\}$, and on the {right} hand side the received signals $\mathsf{Y}_\mathsf{p}$ and $\mathsf{Y}_\mathsf{c}$, which are vectors of normalized length $\max\{1,\alpha\}$ and are the sum of a certain down shifted version of the transmitted vectors. After the down-shift operation, the top part of a vector would be populated by zero; we do not represent these zeros and instead leave an empty space in order not to clutter the figure. Note that the bits received at the same level at a node must be summed modulo-two.}  
The bit vectors $(b_{1},b_{2},b_{3})$ are from PTx to PRx, and the bit vector $b_{4}$ from CTx to CRx. Since the CTx can only either receive or transmit at any point in time, we divide the transmission into two phases.
{\em Phase~1}: for a fraction $\gamma\in[0,1]$ of the time  CTx listens to the channel 
and PTx sends $(b_{1},b_{2})$; $b_{1}$ is decoded by PRx and $b_{2}$ is decoded only at CTx.
{\em Phase~2}: for the remaining fraction $1-\gamma$ of the time CTx transmits; PTx sends $(b_{2},b_{3})$ and CTx sends $(b_{2},b_{4})$ -- notice that PTx and CTx cooperate in sending $b_{2}$, which 
hence is a {\em cooperative message}. The vectors $b_{i}, i\in\{3,4\},$ are split into a {\em common message} ($b_{ic}$), decoded also at the non-intended receiver, and a {\em private message} ($b_{ip}$), treated as noise at the non-intended receiver.

More specifically, in Phase~1 in Fig.~\ref{fig:firstphase} CTx listens to the channel and PTx sends the vector $[b_{1}, b_{2}]$, where $b_{1}$ has normalized {(by the direct link gain $n_\mathsf{d}$)} length $1$ and $b_{2}$ has normalized length $\beta-1$ {(for a total normalized length of $1+(\beta-1)=\beta = \max\{1,\alpha,\beta\}$)}. Hence, over a fraction $\gamma$ of the transmission time, CTx receives $\gamma(\beta-1)$ $b_{2}$-bits that PRx has not received yet. 
In Phase~2, CTx assists PTx to deliver these $b_{2}$-bits to PRx in either of the two following cooperation modes:
(i) CTx relays the $b_{2}$-bits to PRx on behalf of PTx by spending some of its own resources,
(ii) CTx treats the $b_{2}$-bits as a `state non-causally known at the transmitter but unknown at the receiver' and precodes its transmitted signal against it.

Phase~2 in Fig.~\ref{fig:secondphase8}: CTx sends the vector $[b_{4c}, 0, b_{4p}+b_{2}]$, whose components have normalized lengths $2\alpha-1$, $1-\alpha$ and $1-\alpha$, respectively. In the LDA, the linear combination $b_{4p}+b_{2}$ can be thought of as pre-coding the signal $b_{4p}$ against the interference caused by $b_{2}$. PTx sends the vector $[b_{3c}, b_{2}, 0,b_{3p}]$, whose components have normalized lengths $2\alpha-1$, $2-3\alpha$, $2\alpha-1$ and $1-\alpha$ (with an abuse of notation, here $b_{2}$ indicates the bits that have been received in Phase~1 {at CTx}), respectively. 
CRx successively decodes $b_{4c}, b_{3c}, b_{4p}$ in this order, while PRx successively decodes $b_{3c}, b_{2}, b_{4c}, b_{3p}$ in this order. Notice that CRx does not experience interference from $b_{2}$ when decoding $b_{4p}$ (recall that on GF(2) $1+1=0+0=0$).
The achievable rates are
 ${R}_\mathsf{p}/n_\mathsf{d}=\gamma \cdot 1+ (1-\gamma) \cdot (2-2\alpha)$ and
 ${R}_\mathsf{c}/n_\mathsf{d}=\gamma \cdot 0+ (1-\gamma) \cdot \alpha$, thus giving the sum-rate
 $\left({R}_\mathsf{p}+{R}_\mathsf{c}\right)^{\rm(IB)}/n_\mathsf{d}=\gamma \cdot 1+ (1-\gamma) \cdot (2-\alpha)$. 
This sum-rate is larger than that without cooperation, given by $2\mathsf{d}^{\rm(NoCoop)}=2 \alpha$ \cite{etw}, if
$\gamma \leq \frac{2-3\alpha}{1-\alpha}$.
Next, $\gamma^\star$ in~\eqref{eq:gamma opt sum very weak} is smaller than $\frac{2-3\alpha}{1-\alpha}$ only if
$\beta > 2 \alpha$. Thus, when $\beta \leq 2 \alpha$, it would take too much time for the CTx to learn the message of the PTx and it is therefore better to not cooperate at all.
The last observation gives an intuitive interpretation of why the gDoF in~\eqref{eq:gdof sym final} contains the term $[\beta - 2 \alpha]^+$ for $\alpha \in [1/2,2/3)$: the gDoF without cooperation is improved by  HD unilateral cooperation only when $\beta>2\mathsf{d}^{\rm(NoCoop)} = 2 \alpha$.

A similar reasoning may be done for the case $2-3\alpha>\alpha$, which corresponds to $\alpha \in [0,1/2)$.
For this regime an achievable scheme is given in Figs.~\ref{fig:firstphase} and~\ref{fig:secondphase10}, and the gDoF without cooperation is improved by HD unilateral cooperation only when $\beta > 2\mathsf{d}^{\rm(NoCoop)} = 2-2 \alpha$. The scheme for $\alpha \in [0,1/2)$ is simpler than the one for $\alpha \in [1/2,2/3)$ in that it only involves private messages. In particular, CTx sends the vector $[b_{4p}+b_{2}]$ (i.e., $b_{4p}$ is DPC-ed against $b_{2}$), of normalized length $1$; PTx sends the vector $[b_{2}, b_{3p}, 0]$, whose components have normalized lengths $\alpha$, $1-2\alpha$ and $\alpha$, respectively; CRx decodes $b_{4p}$ interference free because of DPC; PRx decodes $b_{2}$ and $b_{3p}$ in this order; the optimal $\gamma$ is such that the amount of $b_{2}$-bits received by CTx in Phase~1 can be delivered to PRx in Phase~2, that is, $\gamma(\beta-1)=(1-\gamma)\alpha$ thus giving the $\gamma^\star$ in~\eqref{eq:gamma opt sum very weak} for $\alpha<1/2$; the achievable sum-rate is $\left({R}_\mathsf{p}+{R}_\mathsf{c}\right)^{\rm(IB)}/n_\mathsf{d}=\gamma \cdot 1+ (1-\gamma) \cdot (2-\alpha)$.

{\bf Very strong interference regime: $\alpha\geq 2$.}
Without cooperation, i.e., $\beta=0$, the tightest upper bound in this regime is~\eqref{eq:gdof sym cs}~\cite{etw}.
For a general $\beta > 0$, the bound in~\eqref{eq:gdof sym cs} is optimized by $\gamma=0$, which is equivalent to no-cooperation, whenever
$\max\{1,\beta\} \leq 2=2\mathsf{d}^{\rm(NoCoop)}$, which is equivalent to
$\beta \leq 2\mathsf{d}^{\rm(NoCoop)}$.
Again we see that  HD unilateral cooperation is beneficial in terms of gDoF only when $\beta$ is larger than the sum-gDoF without cooperation. Here the optimal $\gamma$ is obtained by equating the bounds in~\eqref{eq:gdof sym cs} and~\eqref{eq:gdof sym dt} and given by
\begin{align}
\gamma^\star = \frac{\alpha-2}{\beta+\alpha-3}.
\label{eq:gamma opt sum very strong}
\end{align}

To see why the optimal $\gamma$ is given by~\eqref{eq:gamma opt sum very strong}, we again first analyze the LDA.
Phase~1 is the same as in Fig.~\ref{fig:firstphase}.
In Phase~2 / Fig. \ref{fig:secondphase3}, CTx sends $[b_{4c}, b_{2}, 0]$, whose components have normalized lengths  $1$, $\alpha-2$, and $1$  (here $b_{2}$ indicates again, with an abuse of notation, the bits that have been received in Phase~1 at CTx), respectively.  PTx sends $[b_{3c},0]$, whose components have normalized lengths  $1$ and $\alpha-1$, respectively. CRx successively decodes $b_{3c}, b_{4c}$ in this order. PRx successively decodes $b_{4c}, b_{2}, b_{3c}$ in this order.
The achievable rates are
 ${R}_\mathsf{p}/n_\mathsf{d}=\gamma \cdot 1+ (1-\gamma) \cdot (\alpha-1)$ and
 ${R}_\mathsf{c}/n_\mathsf{d}=\gamma \cdot 0+ (1-\gamma) \cdot 1$, giving a sum-rate of
 $\left({R}_\mathsf{p}+{R}_\mathsf{c}\right)^{\rm(IB)}/n_\mathsf{d}=\gamma \cdot 1+ (1-\gamma) \cdot \alpha$. 
This sum-rate is larger than that without cooperation, given by $2\mathsf{d}^{\rm(NoCoop)}=2$ \cite{etw}, if
$
\gamma \leq \frac{\alpha-2}{\alpha-1}$.
Next, $\gamma^\star$ in~\eqref{eq:gamma opt sum very strong} is smaller than $\frac{\alpha-2}{\alpha-1}$ only if
$
\beta > 2$.
Again, the interpretation is that, if $\beta \leq  2$, it takes too long to transfer bits from PTx to CTx and hence it is preferable not to cooperate. This last observation gives an intuitive interpretation of why the gDoF in~\eqref{eq:gdof sym final} contains the term $[\beta - 2 ]^+$ for $\alpha \in [2, \infty)$: the gDoF without cooperation is improved only when $\beta > 2\mathsf{d}^{\rm(NoCoop)} = 2$. 

{\bf Moderately weak and strong interference regimes:}
For $\alpha\in[2/3,2)$ and without cooperation $\beta=0$ the upper bound in~\eqref{eq:gdof sym dt} is the tightest~\cite{kramer}.
The bound in~\eqref{eq:gdof sym dt} is always optimized by $\gamma=0$, which is equivalent to the case of no-cooperation. Hence, in this regime it is always gDoF-optimal to operate the channel as a non-cooperative IC
and HD unilateral cooperation does not help in managing interference. 
It is very surprising that in this regime, no matter how strong the cooperation link is, unilateral causal cooperation cannot beat the performance of the non-cooperative system. In other words, $\mathsf{d}^{\rm(NoCoop)}=\mathsf{d}^{\rm(SYM)}=\mathsf{d}^{\rm(Ideal)}$ for $\alpha\in[2/3,2)$. 
For $\alpha\in[2/3,1)$, an optimal scheme for the LDA only uses $b_{3c},b_{3p}$ at PTx and $b_{4c},b_{4p}$ at CTx \cite{bresler};
for $\alpha\in[1,2)$, $b_{3c}$ at PTx and $b_{4c}$ at CTx, both of normalized length $\alpha/2$, are optimal \cite{bresler}.

{\bf From the LDA to the AWGN:}
In Appendix~\ref{app:ach}, we show how the LDA schemes in Fig.~\ref{fig:fig5}
can be `translated' into schemes for the G-HD-CCIC that are to within a constant gap from the upper bound in \eqref{eq:upupup}. The `translation' is as follows:
(i) the different pieces of information conveyed through the $b$-vectors  in the LDA correspond to independent Gaussian codewords which are summed together and sent through the G-HD-CCIC; 
(ii) the position, from top to bottom in Fig.~\ref{fig:fig5}, of a $b$-vector within the transmit signal vector in the LDA corresponds to the transmit power of the corresponding Gaussian codeword in the G-HD-CCIC; the higher the position of the $b$-vector, the larger the power of the corresponding Gaussian codeword;
(iii) the length of a $b$-vector in the LDA corresponds to the rate of the corresponding Gaussian codeword in the G-HD-CCIC; the longer the $b$-vector, the higher the rate of the corresponding Gaussian codeword;
(iv) the transmission of the sum of two $b$-vectors in the LDA corresponds to a Gaussian codeword being DPC-precoded against known interference in G-HD-CCIC;
(v) at the receiver side, stripping decoding is used with the Gaussian codeword corresponding to the top-most not-yet-decoded $b$-vector in the received LDA signal being decoded while treating the other signals as noise. 
Therefore, the LDA schemes in Fig.~\ref{fig:fig5} tell us exactly: (a) how many Gaussian codewords must be superposed, with which power and at what rate, (b) if DPC is needed and if so against which interfering codeword, and (c) the decoding order at the receivers.
With this, the achievable scheme is completely specified and the achievable 
rate can be computed.

\begin{figure}
\centering
\includegraphics[width=\columnwidth]{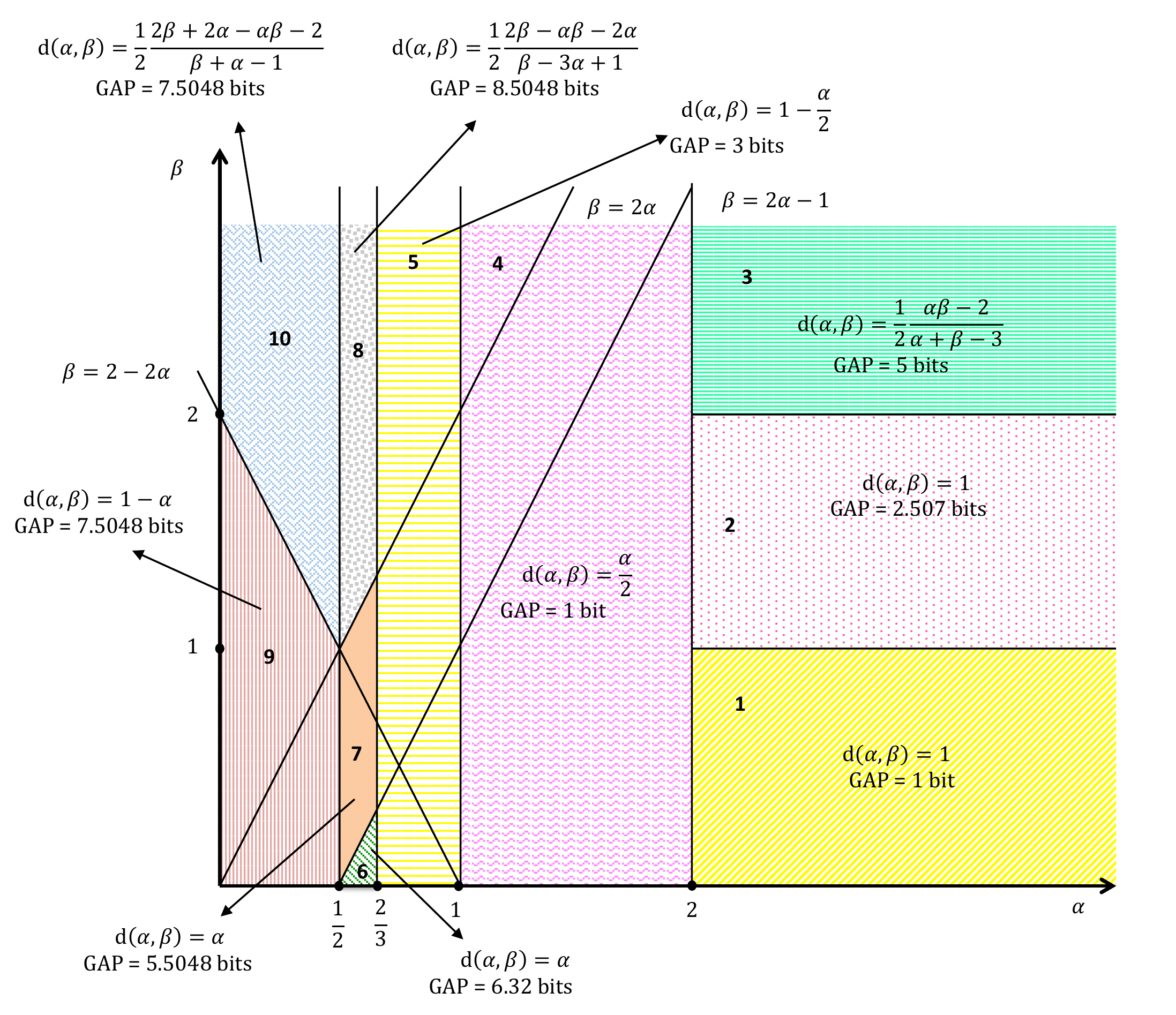}
\caption{gDoF and constant gap for the interference-symmetric G-HD-CCIC.}
\label{fig:fig3}
\vspace{-4mm}
\end{figure}

{\bf Comparisons:}
Fig. \ref{fig:fig3} shows the optimal gDoF and the gap for the interference-symmetric G-HD-CCIC. The whole set of parameters has been partitioned into multiple sub-regions depending upon different levels of cooperation ($\beta$) and interference ($\alpha$) strengths. These regimes are numbered from~1 to~10, {each discussed in} Appendix~\ref{sec:constGap}.  We end this analysis with few comments:
(i) everywhere, except in regions 3, 8 and 10 in Fig.~\ref{fig:fig3}, HD unilateral cooperation might not be worth implementing since the same gDoF is achieved without cooperation;
(ii) the symmetric G-HD-CCIC attains the same gDoF of the non-causal G-CIC in regions 4 and 5 in Fig.~\ref{fig:fig3}. Thus, in these two regions, the performance of the system, in terms of gDoF, is not worsened by allowing causal learning at CTx;
(iii) in regions 1, 4, 5 and 6 of Fig.~\ref{fig:fig3} the gDoF equals that of the equivalent FD channel \cite{cardoneJ1} and is equal to the non-cooperative case. Since the FD channel is an outer bound for the HD channel and no-cooperative strategies are possible under the HD constraint, we conclude that in these regimes the same gap results found for the FD case \cite{cardoneJ1} hold in the case of HD source cooperation.  In this case, gDoF-wise, there is no loss in having a HD CTx compared to a more powerful FD CTx;
(iv) all the achievable schemes use successive decoding at the receivers, which, in practice, is simpler than joint decoding. Thus our proposed schemes, which are optimal to within a constant gap, may be used as guidelines to deploy practical cognitive radio systems;
(v) the gap computed in this work is around $2$~bits larger than that computed in the corresponding FD case \cite{cardoneJ1}. Possible ways to reduce the gap may be:
 (a) apply joint decoding at the receivers;
 (b) develop block Markov coding schemes instead of taking inspiration by the LDA;
 (c) develop achievable strategies that exploit the randomness into the switch to convey further useful information;
 (d) develop tighter upper bounds than those used in this work, especially for $\alpha \leq  2/3$ where the gap is quite large.

\begin{figure}[h]
\centering
\includegraphics[width=\columnwidth]{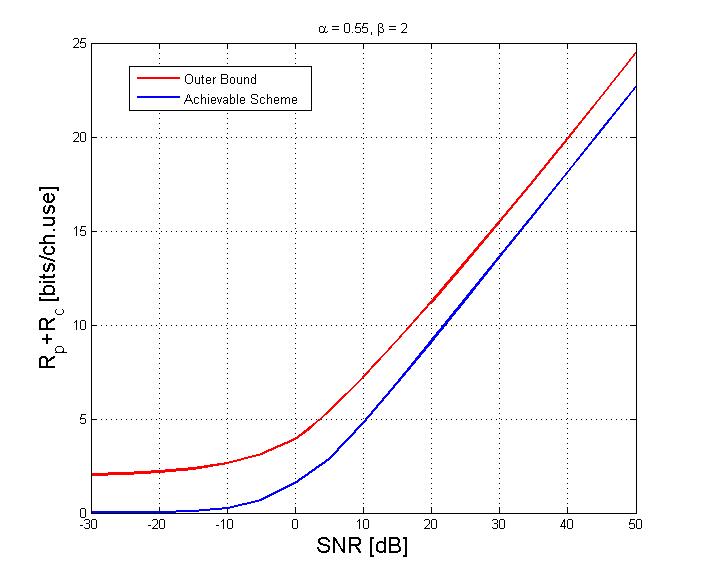}
\caption{Numerical evaluation of the gap for the symmetric G-HD-CCIC with $\alpha=0.55$ and $\beta=2$ (Region 8 in Fig.~\ref{fig:fig3}).}
\label{fig:sim}
\vspace{-4mm}
\end{figure}

{\bf On numerical evaluation of the gap:}
The gap in Theorem \ref{thm:dofuniHDIFC} is pessimistic and it is due to the crude bounding of the upper and lower bounds, which seems to be necessary to obtain expressions that can be easily handled and compared analytically. In order to illustrate this point, in Fig.~\ref{fig:sim} we numerically evaluate the outer bound in \eqref{eq:upupup} and the lower bound obtained from the scheme in Appendix \ref{sec:ach8sym} when the channel parameters fall into region 8 in Fig.~\ref{fig:fig3}, where the gap is the largest. By numerically optimizing all the optimization variables, i.e., power splits, correlation coefficients, fraction of time the CTx listens, we observe from Fig.~\ref{fig:sim} that the gap is of around $2.4$~bits, i.e., more than $6$~bits less than the analytical one. Although we can claim this gap reduction only for the simulated set of channel gains, we believe that this is a more general result.

\section{The $Z$-channel}
\label{sec:asymm Z}
In this section we analyze one of the two interference-asymmetric G-HD-CCIC models: the $Z$-channel.
With $\mathsf{I}_\mathsf{p}=0$ we obtain the $Z$-channel, for which we have
\begin{theorem}
\label{thm:dofuniHDIFC-Z}
The sum-capacity upper bound in \eqref{eq:upupup} is achievable to within \gapz \ bits  regardless of the actual value of the channel parameters for the $Z$-channel.
Therefore, the gDoF can be obtained from~\eqref{eq:upupup} {with $\mathsf{I}_\mathsf{p}=0$} and equals
\begin{align}
\mathsf{d}^{\rm(Z)}
& =  \frac{1}{2}\max_{\gamma\in[0,1]} \min 
  \left \{ \gamma \max \left \{1,\beta \right \} + \left( 1-\gamma \right) 2, \right. \nonumber
\\& \quad \left.  \gamma  + \left( 1 -  \gamma \right ) \left( \max \left \{ 1,\alpha \right \} +  \left [ 1 - \alpha \right ]^+ \right )\right \} \nonumber
\\&=\left\{\begin{array}{ll}
\max\left\{1-\frac{1}{2}\alpha,\frac{1}{2}\alpha \right\} & \alpha\in[0,2) \\
1+\frac{1}{2}\frac{[\beta-2]^+ \ (\alpha-2)}{\beta + \alpha-3} & \alpha\in[2,\infty) \\
\end{array}.\right.
\label{eq:gdof z final}
\end{align}
\end{theorem}
\begin{IEEEproof}
The details of the proof can be found in Appendix~\ref{sec:constGapz}. 
\end{IEEEproof}

For future reference, for the Z-channel
$\mathsf{d}^{\rm(NoCoop)} = \min\{1, \max\{1-\alpha/2,\alpha/2\}\}
\leq \mathsf{d}^{\rm(Z)} \leq
\mathsf{d}^{\rm(Ideal)}  =\max\{1-\alpha/2,\alpha/2\}$
hence cooperation can only improve the gDoF in very strong interference, i.e., $\alpha > 2$.
The interpretation of the gDoF in~\eqref{eq:gdof z final} is similar to that of the interference-symmetric case in~\eqref{eq:gdof sym final}.  In particular, if the channel has weak or strong interference, i.e., $\alpha \leq 2$, the gDoF is the same as for the non-cooperative $Z$-channel \cite{sason}; in this regime it might not be worth to engage in unilateral cooperation. In very strong interference, i.e., $\alpha > 2$, unilateral cooperation gives larger gDoF than in the case of no-cooperation only when $\beta > 2\mathsf{d}^{\rm(NoCoop)} = 2$. 
An achievable scheme for the LDA in this regime is exactly the same developed for the corresponding interference-symmetric channel in Figs. \ref{fig:firstphase} and~\ref{fig:secondphase3}, with the only difference that now the signal $\mathsf{X}_\mathsf{p}[2]$ is not received at $\mathsf{Y}_\mathsf{c}[2]$ since ${\mathsf {I}}_\mathsf{p}=0$.

\begin{figure}
\centering
\includegraphics[width=\columnwidth]{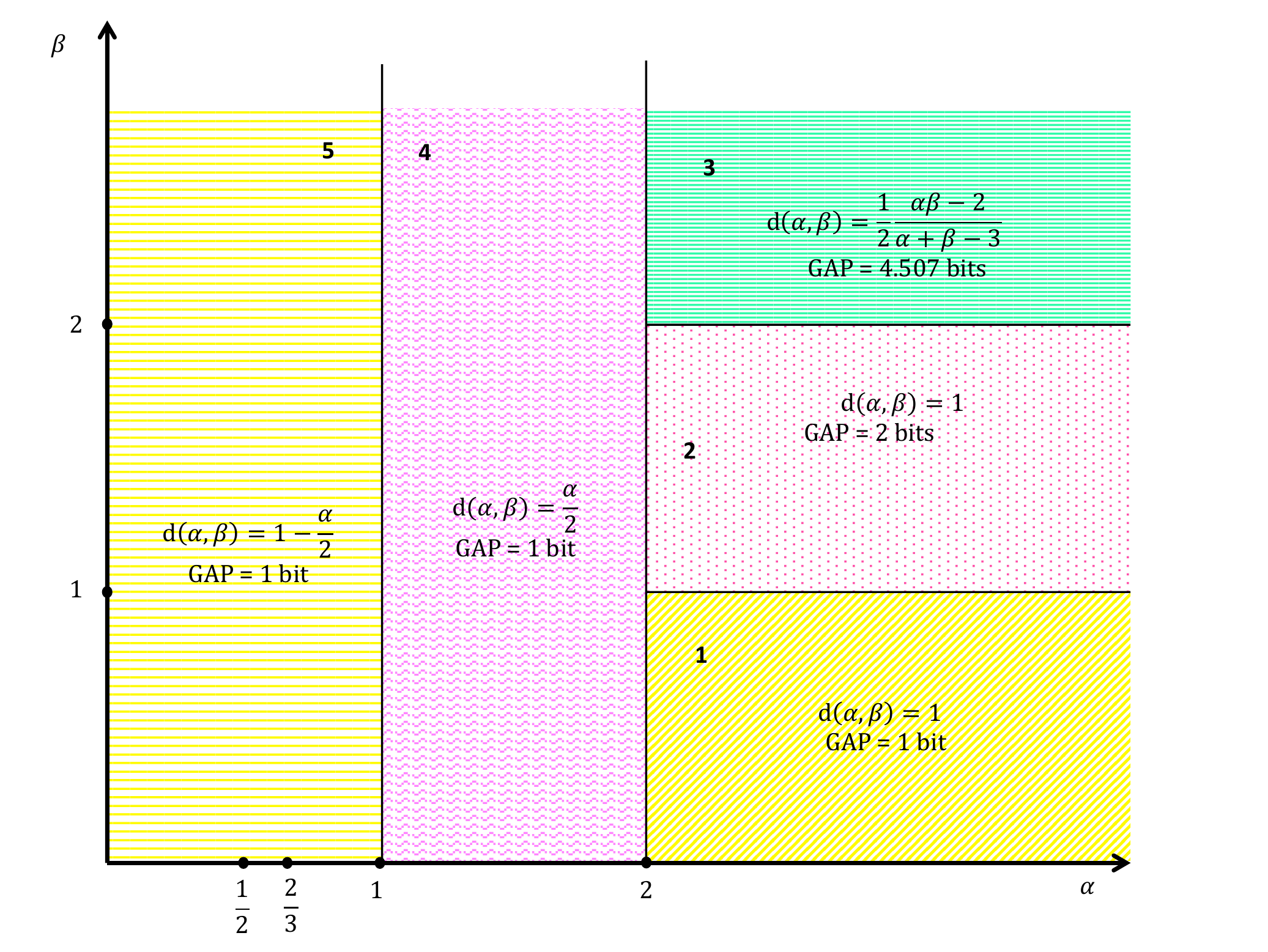}
\caption{gDoF and constant gap for the interference-asymmetric Z G-HD-CCIC.}
\label{fig:fig4}
\vspace{-4mm}
\end{figure}

Fig. \ref{fig:fig4} shows the optimal gDoF and the gap for the $Z$-channel. The whole set of parameters has been partitioned into multiple sub-regions depending upon different levels of cooperation ($\beta$) and interference ($\alpha$) strengths. These regimes are numbered from~1 to~5 and the details for the gap computation appear in Appendix~\ref{sec:constGapz}. 
We conclude with few comments:
(i) 
in regions 1, 4 and 5 of Fig. \ref{fig:fig4} the gDoF of the HD channel is as that in FD \cite{cardoneJ1}
and so the same gap results found for the FD case hold in the HD case. Moreover, in region 2 in Fig. \ref{fig:fig4} the gDoF equals that of the non-cooperative $Z$-channel. Hence, in regions 1, 2, 4 and 5 cooperation might not be worth implementing since the same gDoF is attained without cooperation;
(ii) the $Z$-channel achieves the same gDoF of the non-causal cognitive $Z$-channel everywhere except in $\alpha>2$ (regions 1, 2 and 3 in Fig.~\ref{fig:fig4}), i.e., 
for $\alpha \leq 2$, causal cognition attains the ultimate performance of the ideal non-causal cognitive radio $Z$-channel;
(iii) by comparing Figs.~\ref{fig:fig3} and~\ref{fig:fig4}, we observe that the gDoF of the $Z$-channel is always greater than or equal to that of the interference-symmetric channel. This is 
because the PTx does not cooperate in sending the message of CTx, i.e.,
by removing the link between PTx and CRx we rid CRx of only interfering signals. 
We observe that the $Z$-channel  outperforms the interference-symmetric G-HD-CCIC when $0 \leq \alpha \leq  2/3$.

\section{The $S$-channel}
\label{sec:asymm S}
In this section we analyze the other interference-asymmetric G-HD-CCIC model: the $S$-channel.
With $\mathsf{I}_\mathsf{c}=0$ we obtain the $S$-channel, for which we have
\begin{theorem}
\label{thm:dofuniHDIFC-S}
The sum-capacity upper bound in \eqref{eq:upupup} is achievable to within \gaps \ bits  regardless of the actual value of the channel parameters for the $S$-channel.
Therefore, the gDoF can be obtained from~\eqref{eq:upupup} {with $\mathsf{I}_\mathsf{c}=0$} and equals
\begin{align}
&\mathsf{d}^{\rm(S)} 
 =  \frac{1}{2}\max_{\gamma\in[0,1]} \min 
  \left \{ \gamma + 2\left( 1 - \gamma \right), \right. \nonumber
\\& \left. \gamma \max \left \{ \beta,\alpha,1 \right \} +\left( 1 -  \gamma \right ) \left( \max \left \{ 1,\alpha \right \} +  \left [ 1 - \alpha \right ]^+ \right )\right \} \nonumber 
\\&=
\left\{\begin{array}{ll}
1-\frac{1}{2}\alpha+\frac{1}{2} \frac{\alpha \left [ \alpha+\beta-2 \right ]^+}{\beta+\alpha-1}    & \alpha\in[0,1) \\
  \frac{1}{2}\alpha+\frac{1}{2} \frac{(2-\alpha) \left [ \beta-\alpha \right ]^+}{\beta-\alpha+1}  & \alpha\in[1,2) \\
1 & \alpha\in[2,\infty) \\
\end{array}.\right.
\label{eq:gdof s final}
\end{align}
\end{theorem}
\begin{IEEEproof}
The details of the proof can be found in Appendix~\ref{sec:constGaps}. 
\end{IEEEproof}

For future reference, for the $S$-channel
$\mathsf{d}^{\rm(NoCoop)} = \min\{1, \max\{1-\alpha/2,\alpha/2\}\}
\leq \mathsf{d}^{\rm(S)}  \leq 
\mathsf{d}^{\rm(Ideal)}  =1$
hence cooperation can improve performance only if the channel is not in very strong interference, i.e.,  $\alpha < 2$. {    It is interesting to notice the different behavior of the $Z$- and $S$-channel: for the $Z$-channel HD unilateral cooperation is useful only in very strong interference, while for the $S$-channel only when not in very strong interference.}
Also in this case the interpretation of the gDoF in~\eqref{eq:gdof s final} is similar to that of the interference-symmetric case in~\eqref{eq:gdof sym final}.  In particular, if the channel has very strong interference, i.e., $\alpha > 2$, the gDoF is the same as for the non-cooperative $S$-channel \cite{sason}; in this regime it might not be worth to engage in unilateral cooperation. In weak and strong interference, i.e., $\alpha \leq 2$, unilateral cooperation gives larger gDoF than in the case of no-cooperation only when $\beta > 2\mathsf{d}^{\rm(NoCoop)} = 2\max\{1-\alpha/2,\alpha/2\}$. A representation of the LDA schemes used for $\alpha < 2$ and $\beta > 2\max\{1-\alpha/2,\alpha/2\}$ is given in Figs.~\ref{fig:firstphase},~\ref{fig:secondphase5S} and~\ref{fig:secondphase4S}, which can be interpreted as done for the interference-symmetric case in Fig. \ref{fig:fig5}.

\begin{figure}[h]
\centering
\includegraphics[width=\columnwidth]{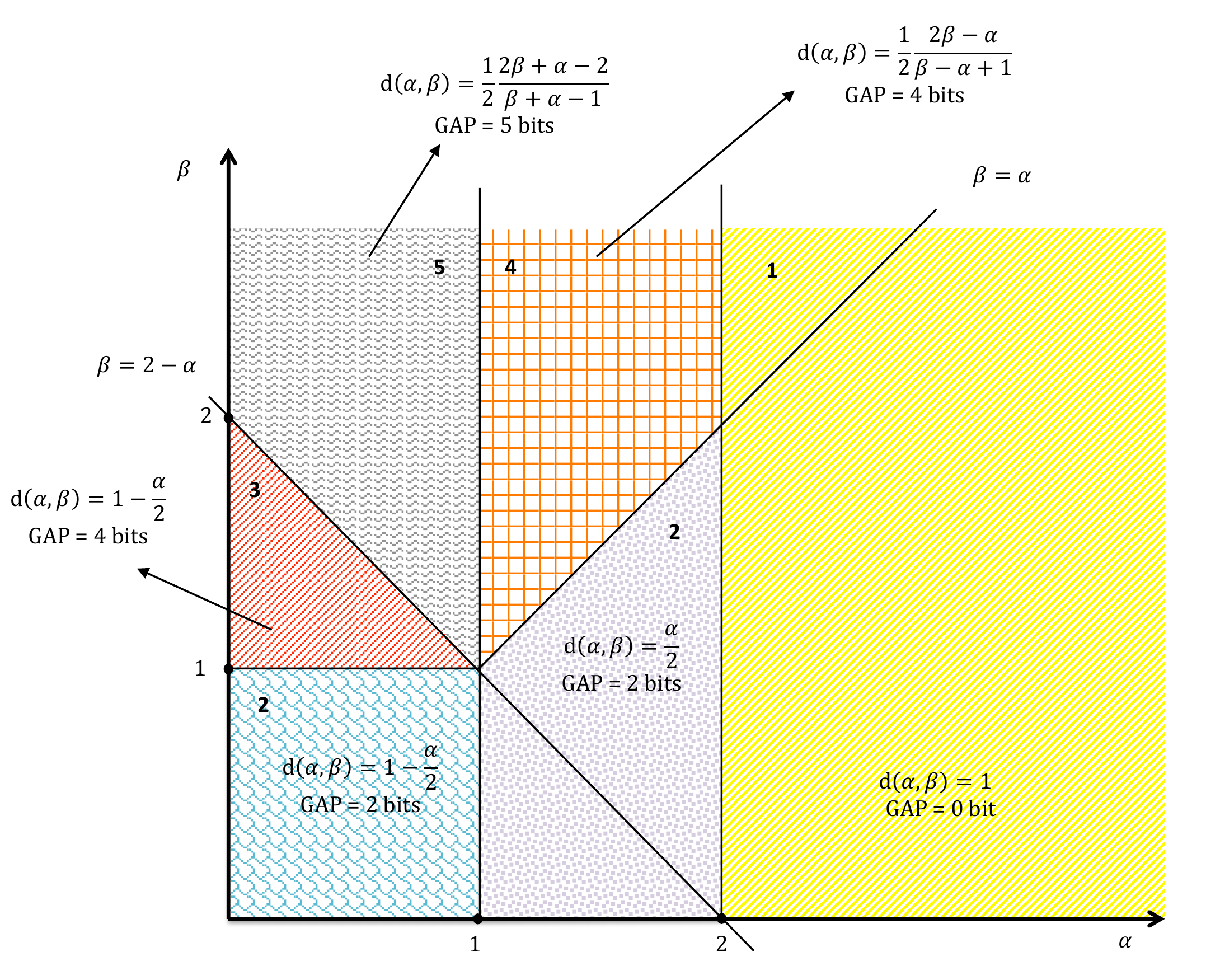}
\caption{gDoF and constant gap for the interference-asymmetric S G-HD-CCIC.}
\label{fig:fig6}
\vspace{-4mm}
\end{figure}

Fig. \ref{fig:fig6} shows the optimal gDoF and the gap for the $S$-channel. The whole set of parameters has been partitioned into multiple sub-regions depending upon different levels of cooperation ($\beta$) and interference ($\alpha$) strengths. The details for the gap computation appear in Appendix~\ref{sec:constGaps}. 
We conclude the analysis of the $S$-channel with few comments:
(i) there are some regions (1 and 2 in Fig. \ref{fig:fig6}) in which the gDoF of the HD channel is as that in FD. In these regions, the same additive gap results found for the FD case \cite{cardoneJ1} hold in HD. Moreover, in region 3 in Fig. \ref{fig:fig6} the gDoF equals that of the non-cooperative $S$-channel;
(ii) the $S$-channel achieves the same gDoF of the non-causal cognitive $S$-channel, i.e., $\mathsf{d}=1$, for $\alpha\geq 2$ (region 1 in Fig.~\ref{fig:fig6}). Thus, in this region the $S$-channel attains the ultimate performance of the ideal non-causal cognitive radio $S$-channel;
(iii) the $S$-channel outperforms the interference-symmetric G-HD-CCIC when either $0  \leq  \alpha \leq   2/3$ or when $\alpha \leq 2$ and $\beta \geq \max \{ 2-\alpha, \alpha \}$ (regions 4 and 5, and parts of regions 2 and 3 in Fig.~\ref{fig:fig6}). On the other hand, the interference-symmetric G-HD-CCIC outperforms the $S$-channel in very strong interference and strong cooperation, i.e., $\min \{\alpha,\beta \} \geq 2$. This is so because, in the very strong interference and cooperation regime, the system performance is enhanced by allowing the CTx to help the PTx to convey the information {to the PRx}, but this is not possible since 
$\mathsf{I}_{\mathsf{c}}=0$;
(iv) when $\alpha \geq 2$ (region 1 in Fig.~\ref{fig:fig6}) we have an exact sum-capacity result, i.e., the gap between the sum-rate outer bound and inner bound is equal to zero.

\section{Conclusions}
\label{sec:Concl}
In this work we studied the Gaussian causal cognitive interference channel where the cognitive source, who helps the primary source, is constrained to operate in half-duplex mode.
From an application standpoint, this model fits future 4G cellular networks with heterogeneous deployments.
We analyzed both the interference-symmetric and interference-asymmetric channels, which correspond to different network deployments. 
We determined, for each topology, the gDoF, the sum-capacity to within a constant gap and the optimal parameters in closed form.
We compared the interference-symmetric and interference-asymmetric models by highlighting the regimes where the gDoF is as that of the classical IC without cooperation, i.e., regimes where cooperation might not be worth implementing, and by identifying the regimes where the system attains the ultimate limits predicted by the ideal non-causal cognitive model, i.e., regimes where the performance is not worsened by allowing causal half-duplex learning at the cognitive source. Extensions to the general channel and to the whole capacity region characterization to within a constant gap are subjects of current investigation.

\section*{Acknowledgment}
The work of D.~Tuninetti was partially funded by NSF under award number 0643954; the contents of this article are solely the responsibility of the authors and do not necessarily represent the official views of the NSF.
Eurecom's research is partially supported by its industrial partners: BMW Group Research \& Technology, IABG, Monaco Telecom, Orange, SAP, SFR, ST Microelectronics, Swisscom and Symantec. The research
carried out at Eurecom leading to these results has received
funding from the EU Celtic+ Framework Program Project
SPECTRA.
The research at IMC leading to these results has received funding from the EU FP7 project iJOIN, grant agreement $\rm{n}^\circ$ 317941. 

\appendices

\section{}
\subsection{Interference-symmetric G-HD-CCIC}
\label{sec:constGap}
Let $\mathsf{I}_\mathsf{p}=\mathsf{I}_\mathsf{c}=\mathsf{I}$ and $\mathsf{d}^{\rm(SYM)}=\mathsf{d}$ for brevity. We analyze the different regimes in Fig. \ref{fig:fig3}.

{ {\bf{Regime 1: Very Strong Interference 1}}: $\alpha \geq 2, \ \beta \leq 1$}. 
\textit{Parameter Range}: $\mathsf{I}  \geq  \mathsf{S}(1 + \mathsf{S})$ and $\mathsf{C}  \leq  \mathsf{S}$, in which case we have $\mathsf{d}  \leq  1$ (with $\gamma = 0$), as in FD.  
Thus, as in FD, $\mathsf{GAP}  \leq  1$~bit.

{ {\bf{Regime 2: Very Strong Interference 2}}: $\alpha \geq 2, \ 1 < \beta \leq 2$. 
\textit{Parameter Range}: $\mathsf{I}  \geq  \mathsf{S}(1 + \mathsf{S})$ and $\mathsf{S}  <  \mathsf{C}  \leq  \mathsf{S}(\mathsf{S} + 1)$, in which case $\mathsf{d}  \leq  1$ (with $\gamma = 0$).
\textit{Inner Bound}: classical IC in very strong interference, i.e.,
$\left({R}_\mathsf{p} + {R}_\mathsf{c} \right)^{\rm(IB)}  =  2\log \left(1 + \mathsf{S}  \right)$,
which implies
$\mathsf{d}  =  1$.
\textit{Outer Bound}: in this regime the tightest sum-rate outer bound is given by \eqref{eq:CutSet}, which can be further upper-bounded as
$\left({R}_\mathsf{p}+{R}_\mathsf{c} \right)^{\rm(OB)}  \leq 2.507 +2 \log \left(1+\mathsf{S}  \right)$. Thus,
$\mathsf{GAP}\leq 2.507 \ \rm{bits}$.

{ {\bf{Regime 3: Very Strong Interference 3}}: $\alpha  \geq 2, \ \beta  > 2$. 
\textit{Parameter Range}: $\mathsf{I} \geq \mathsf{S}(1+\mathsf{S})$ and $\mathsf{C} > \mathsf{S}(\mathsf{S}+1)$, where we have
$2\mathsf{d} \leq \max_{\gamma} \min \left \{\gamma \beta  + 2 \left( 1 - \gamma \right),
\gamma +\left( 1-\gamma \right) \alpha \right \}$.
In this expression the first term is increasing in $\gamma$ while the second one is decreasing in $\gamma$. Thus the optimal $\gamma$ can be found by equating the two terms and is given by
$\gamma^\star = \frac{\alpha-2}{\beta+\alpha-3}$ in \eqref{eq:gamma opt sum very strong}, which leads to
$\mathsf{d} \leq \frac{1}{2} \frac{\beta \alpha -2}{\beta + \alpha-3}$.
\textit{Inner Bound}: The achievable scheme in Appendix \ref{sec:ach3sym}, with 
$\gamma^\prime  =  \frac{x}{\log \left( 1 + \frac{\mathsf{C}}{1 + \mathsf{S}} \right) + x}$ $\stackrel{\mathsf{SNR} \gg 1}{\to} \gamma^\star$ for $x := \log \left( 1 + \frac{\mathsf{I}}{(1+\mathsf{S})^2} \right) \stackrel{\mathsf{SNR} \gg 1}{\to} \alpha - 2$ (recall that $\log \left( 1 + \frac{\mathsf{C}}{1+\mathsf{S}} \right)  \stackrel{\mathsf{SNR} \gg 1}{\to} \beta - 1 $), gives \eqref{eq:ach1a} at the top of next page,
\begin{figure*}
\begin{align}
&\left(\mathsf{R}_\mathsf{p} + \mathsf{R}_\mathsf{c}\right)^{\rm(IB)} =\gamma^\prime \log\left(1+\mathsf{S}\right) -\log \left( 1+\frac{\mathsf{S}}{1+\mathsf{S}}\right) +\gamma^\prime \log \left( 1+\frac{\mathsf{C}}{1+\mathsf{S}} \right)+ 2\left( 1-\gamma^\prime \right)\log\left(1+\mathsf{S}\right)
\label{eq:ach1a}
\\&\left(\mathsf{R}_\mathsf{p}+\mathsf{R}_\mathsf{c} \right)^{\rm(IB)} = \gamma^\prime \log \left( 1+\mathsf{S}\right) - \gamma^\prime \log \left( 1+\frac{\mathsf{S}}{1+\mathsf{S}} \right)+\gamma^\prime \log \left ( 1+\frac{\mathsf{C}}{1+\mathsf{S}} \right ) + 2\left( 1-\gamma^\prime \right) \log \left( 1+\mathsf{I}+\frac{\mathsf{S}}{1+\mathsf{I}} \right) \nonumber
\\& \qquad -\left( 1-\gamma^\prime \right) \left(2\log \left( 1+\frac{\mathsf{I}}{1+\mathsf{I}} \right)-\log \left( 1+\frac{\mathsf{S}}{1+\mathsf{I}}+\frac{\mathsf{I}}{1+\mathsf{I}} \right) \right)  -\left( 1-\gamma^\prime \right) \log \left( 1+ \frac{\mathsf{S}\mathsf{I}+\mathsf{I}+\mathsf{I}^2}{(1+\mathsf{I})^2} + \frac{\mathsf{S}}{1+\mathsf{I}}\right) 
\label{eq:eqach3a}
\\ &\left(\mathsf{R}_\mathsf{p}+\mathsf{R}_\mathsf{c} \right)^{\rm(IB)} = \gamma^\prime \log \left( 1+\mathsf{S}\right) - \gamma^\prime \log \left( 1+\frac{\mathsf{S}}{1+\mathsf{S}} \right)+\gamma^\prime \log \left ( 1+\frac{\mathsf{C}}{1+\mathsf{S}} \right ) -\left( 1-\gamma^\prime \right) \log \left( 1+\mathsf{I}\right)\nonumber
\\& \qquad + \left( 1-\gamma^\prime \right) \log \left( 1+\mathsf{I}+\frac{\mathsf{S}}{1+\mathsf{I}} \right)  -\left( 1-\gamma^\prime \right) \log \left( 1+\frac{\mathsf{I}}{1+\mathsf{I}} \right) +\left( 1-\gamma^\prime \right) \log \left( 1+\frac{\mathsf{I}}{1+\mathsf{I}}+\mathsf{S} \right)
\label{eq:eqach2a}
\\ &\left(\mathsf{R}_\mathsf{p}+\mathsf{R}_\mathsf{c} \right)^{\rm(IB)}
 = \log \left( 1+\mathsf{S}\right) - \gamma^\prime \log \left( 1+\frac{\mathsf{S}}{1+\mathsf{S}} \right)
+\gamma^\prime \log \left( 1+\frac{\mathsf{C}}{1+\mathsf{S}} \right) \nonumber
 \\& \qquad +\left( 1-\gamma^\prime \right) \log \left( 1+\mathsf{S}+\mathsf{I} \right)
-\left( 1-\gamma^\prime \right) \log \left( 1+\frac{\mathsf{S}\mathsf{I}}{1+\mathsf{I}}+\mathsf{S} \right)
\label{eq:ach S regime 4}
\\ &\left(\mathsf{R}_\mathsf{p}+\mathsf{R}_\mathsf{c} \right)^{\rm(IB)}
= \gamma^\prime\log \left( 1+\mathsf{S}\right) - \gamma^\prime \log \left( 1+\frac{\mathsf{S}}{1+\mathsf{S}} \right)\nonumber
+\gamma^\prime \log \left( 1+\frac{\mathsf{C}}{1+\mathsf{S}} \right)+\left( 1-\gamma^\prime \right) \log \left( 1+\frac{\mathsf{S}}{1+\mathsf{I}}\right)
\nonumber
\\&  \qquad +\left( 1 - \gamma^\prime \right) \log \left( 1 + \mathsf{S} + \frac{\mathsf{I}}{1+\mathsf{I}}\right) -\left( 1 - \gamma^\prime \right)  \log \left( 1 + \frac{\mathsf{I}}{1+\mathsf{I}} \right)
\label{eq:ach S regime 5}
\end{align}
\vspace{-4mm}
\end{figure*}
which implies
$2\mathsf{d}  =  
 \frac{\alpha - 2}{\beta + \alpha - 3}[1 - 0 + (\beta - 1)]
 +  \frac{\beta - 1 }{\beta + \alpha - 3}[1 + 1 - 0]
 = \frac{\alpha \beta - 2}{\alpha + \beta - 3}$.
\textit{Outer Bound}: 
\eqref{eq:CutSet} and \eqref{eq:Tuninetti} are the tightest outer bounds.
With \eqref{eq:CutSet},
$\mathsf{GAP} \leq  3.507 \ \rm{bits}$;
with 
\eqref{eq:Tuninetti}, $\mathsf{GAP}  \leq  5 \ \rm{bits}$.
Thus,
$\mathsf{GAP}  \leq   5 \ \rm{bits}$.

{ {\bf{Regime 4: Strong Interference}}: $1 \leq  \alpha < 2$.
\textit{Parameter Range}: $\mathsf{S} \leq \mathsf{I} < \mathsf{S}(1+\mathsf{S})$, in which case 
$\mathsf{d}  \leq \frac{\alpha}{2}$ (with $\gamma = 0$). 
The gDoF upper bound coincides with that in FD \cite{cardoneJ1}, so
$\mathsf{GAP}  \leq 1$~bit.

{ {\bf{Regime 5: Moderately Weak Interference}}: $2/3 \leq \alpha < 1$.
\textit{Parameter Range}: $\mathsf{I} <\mathsf{S}$ and $\mathsf{S}(\mathsf{S} + 1)  \leq  \mathsf{I}(\mathsf{I} + 1)^2$, 
in which case 
$\mathsf{d}  \leq  1 - \frac{\alpha}{2}$ (with $\gamma = 0$). 
The gDoF upper bound coincides with that in FD \cite{cardoneJ1}. Thus, as in FD,
$\mathsf{GAP}  \leq  3 \ \rm{bits}$.

{ {\bf{Regime 6: Weak Interference 1}}: $1/2 \leq \alpha <  2/3, \ \beta \leq 2\alpha-1$.
\textit{Parameter Range}: $\mathsf{S}(\mathsf{S} + 1)  >  \mathsf{I}(\mathsf{I} + 1)^2$, $ \mathsf{S}  \leq  \mathsf{I} \left(1+\mathsf{I} \right)$ and $\mathsf{C}  \leq  \frac{\mathsf{I}^2}{\mathsf{S}}$, in which case we have
$\mathsf{d}  \leq  \alpha$ (with $\gamma = 0$). The gDoF upper bound coincides with that in FD \cite{cardoneJ1}. Thus, as in FD,
$\mathsf{GAP}  \leq  6.32 \ \rm{bits}$.

{ {\bf{Regime 7: Weak Interference 2}}: $1/2 \leq \alpha <  2/3, \ 2\alpha-1<\beta \leq 2\alpha$.
\textit{Parameter Range}: $\mathsf{S}(\mathsf{S} + 1)   >   \mathsf{I}(\mathsf{I} + 1)^2$, $ \mathsf{S}  \leq  \mathsf{I} \left(1+\mathsf{I} \right)$ and $\frac{\mathsf{I}^2}{\mathsf{S}}   <  \mathsf{C}  \leq  \mathsf{I}^2$, where we have
$\mathsf{d}   \leq   \alpha$  (with $\gamma  =  0$).
\textit{Inner Bound}: classical IC with the power split of \cite{etw}, i.e.,
$\left({R}_\mathsf{p} + {R}_\mathsf{c} \right)^{\rm(IB)}   =   2\log \left(1  +  \mathsf{I}  +  \frac{\mathsf{S}}{\mathsf{I}}  \right)  -  2$,
which implies
$\mathsf{d}=\alpha$.
\textit{Outer Bound}: in this regime the tightest sum-rate outer bound is given by \eqref{eq:Prabh}, that can be further upper bounded as
$\left({R}_\mathsf{p}+{R}_\mathsf{c} \right)^{\rm(OB)} \leq 3.5048 + 2\log \left( 1+ \mathsf{I} + \frac{\mathsf{S}}{\mathsf{I}}\right )$.
Thus, 
$\mathsf{GAP} \leq 5.5048 \ \rm{bits}$.

{ {\bf{Regime 8: Weak Interference 3}}: $1/2 \leq \alpha <  2/3, \ \beta > 2\alpha $.
\textit{Parameter Range}: $\mathsf{S}(\mathsf{S}+1) > \mathsf{I}(\mathsf{I}+1)^2$, $ \mathsf{S}  \leq  \mathsf{I} \left(1+\mathsf{I} \right)$ and $\mathsf{C} >\mathsf{I}^2$, in which case
$2\mathsf{d} \leq \max_{\gamma} \min \left \{\gamma  +\left( 1-\gamma \right) \left( 2-\alpha \right), \gamma  \beta+2 \left( 1-\gamma \right)\alpha \right \}$.
In this expression the first term is decreasing in $\gamma$ while the second term is increasing in $\gamma$. Thus the optimal $\gamma$ can be found by equating the two terms and is given by
$\gamma^\star = \frac{2-3\alpha}{\beta-3\alpha+1}$ in \eqref{eq:gamma opt sum very weak}, which leads to
$\mathsf{d} \leq \frac{1}{2} \frac{2\beta-\alpha \beta -2 \alpha}{\beta -3\alpha+1}$. 
\textit{Inner Bound}: the rate in \eqref{eq:eqach3a} at the top of next page is achievable (Appendix \ref{sec:ach8sym}),
where
$\gamma^\prime = \frac{x}{\log \left( 1+\frac{\mathsf{C}}{1+\mathsf{S}} \right)+x} \stackrel{\mathsf{SNR} \gg 1}{\to} \gamma^\star$ for $x := \log \left( 1+\frac{\mathsf{S}^2}{(1+\mathsf{I})^3+\mathsf{S}+\mathsf{S}\mathsf{I}} \right) \stackrel{\mathsf{SNR} \gg 1}{\to} 2-3\alpha$.
The sum-rate in~\eqref{eq:eqach3a} implies
$2\mathsf{d}
  = 
 \frac{2-3\alpha}{\beta-3\alpha+1} [1-0+(\beta-1)] 
+\frac{\beta-1}  {\beta-3\alpha+1} [2\alpha-0+(1-\alpha)-(1-\alpha)] 
= \frac{2\beta-\alpha \beta -2 \alpha}{\beta -3\alpha+1}$.
\textit{Outer Bound}: in this regime the tightest sum-rate outer bounds are those in \eqref{eq:Tuninetti} and \eqref{eq:Prabh}.
With the bound in \eqref{eq:Tuninetti},
$\mathsf{GAP} \leq  8 \ \rm{bits}$;
with the bound in \eqref{eq:Prabh}, $\mathsf{GAP}  \leq  8.5048 \ \rm{bits}$.
Thus,
$\mathsf{GAP}  \leq   8.5048 \ \rm{bits}$.

{ {\bf{Regime 9: Weak interference 4}}: $\alpha < 1/2, \ \beta \leq 2-2\alpha$.
\textit{Parameter Range}: $\mathsf{I}(\mathsf{I} + 1)  <  \mathsf{S}$ and $\mathsf{C}  \leq   \frac{\mathsf{S}^2}{\mathsf{I}^2}$, in which case we have $\mathsf{d} \leq 1-\alpha$ (with $\gamma=0$).
\textit{Inner Bound}: classical IC, i.e., 
$\left({R}_\mathsf{p}+{R}_\mathsf{c} \right)^{\rm(IB)} = 2\log \left(1+\frac{\mathsf{S}}{1+\mathsf{I}}  \right)$,
which implies
$\mathsf{d}=1-\alpha$.
\textit{Outer Bound}: 
the tightest sum-rate outer bound is given by \eqref{eq:Prabh}, that can be further upper bounded as
$\left({R}_\mathsf{p}+{R}_\mathsf{c} \right)^{\rm(OB)} \leq 3.5048 + 2\log \left( 1+ \mathsf{I} + \frac{\mathsf{S}}{\mathsf{I}}\right )$.
Thus $\mathsf{GAP} \leq 3.5048 + 2 \log \left( \frac{1+ \mathsf{I}}{\mathsf{I}} \right) + 2 \log \left( \frac{\mathsf{I}^2+\mathsf{I}+\mathsf{S}}{1+\mathsf{I}+\mathsf{S}} \right)
 \leq 3.5048 + 4 \log \left( 2 \right )  = 7.5048 \ \rm{bits}$.

{ {\bf{Regime 10: Weak Interference 5}}: $\alpha< 1/2, \ \beta > 2-2\alpha$.
\textit{Parameter Range}: $\mathsf{I}(\mathsf{I}+1) < \mathsf{S}$ and $\mathsf{C} > \frac{\mathsf{S}^2}{\mathsf{I}^2}$, in which case we have $2\mathsf{d} \leq \max_{\gamma} \min \left \{\gamma  +\left( 1-\gamma \right) \left( 2-\alpha \right), \gamma  \beta+2 \left( 1-\gamma \right) \left (1-\alpha \right ) \right \}$.
In this expression the first term is decreasing in $\gamma$ while the second term is increasing in $\gamma$. So the optimal $\gamma$ can be found by equating the two terms and is given by
$\gamma^\star = \frac{\alpha}{\beta+\alpha-1}$ in \eqref{eq:gamma opt sum very weak}, which leads to
$\mathsf{d} \leq\frac{1}{2} \frac{2\beta+2\alpha-\alpha \beta -2}{\beta+\alpha -1}$.
\textit{Inner Bound}: the sum-rate in \eqref{eq:eqach2a} at the top of the page is achievable (Appendix \ref{sec:ach10sym}),
where
$\gamma^\prime= \frac{x}{\log \left( 1+\frac{\mathsf{C}}{1+\mathsf{S}} \right)+x} \stackrel{\mathsf{SNR} \gg 1}{\to} \gamma^\star$ for $x := \log \left( 1+\frac{\mathsf{S}\mathsf{I}}{(1+\mathsf{I})^2+\mathsf{S}} \right) \stackrel{\mathsf{SNR} \gg 1}{\to} \alpha$.
The sum-rate in~\eqref{eq:eqach2a} implies
$2\mathsf{d}
  =\frac{\alpha} {\beta+\alpha-1}[1-0+(\beta-1)]
   +\frac{\beta-1}{\beta+\alpha-1}[- \alpha + (1-\alpha) + 1-0 ]
= \frac{2\beta+2\alpha-\alpha \beta -2}{\beta +\alpha-1}$.
\textit{Outer Bound}: \eqref{eq:Tuninetti} and \eqref{eq:Prabh} are the tightest.
With \eqref{eq:Tuninetti},
$\mathsf{GAP} \leq  6 \ \rm{bits}$;
with \eqref{eq:Prabh}, $\mathsf{GAP}  \leq  7.5048 \ \rm{bits}$.
Thus,
$\mathsf{GAP}  \leq  7.5048 \ \rm{bits}$.

\subsection{$Z$-channel}
\label{sec:constGapz}
Let $\mathsf{I}_\mathsf{c}=\mathsf{I}$, $\mathsf{I}_\mathsf{p}=0$ and $\mathsf{d}^{\rm(Z)}=\mathsf{d}$ for brevity. We consider two different regimes.

{\bf{Regime 1: Strong and Very Strong Interference: $\alpha \geq 1$}}. 
The analysis is similar to that of the interference-symmetric G-HD-CCIC in the same regime (same inner and outer bounds with $\mathsf{I}_\mathsf{p}=0$)
and the gap is at most $4.507 \ \rm{bits}$. 
In regions 1 and 4 in Fig. \ref{fig:fig4}, then, the gDoF upper bound coincides with that in FD~\cite{cardoneJ1}. Thus in these two regions, as in FD, $\mathsf{GAP}\leq 1$~bit.

{\bf{Regime 2: Weak Interference: $\alpha < 1$}},
in which case 
$\mathsf{d} \leq 1-\frac{\alpha}{2}$ (with $\gamma=0$).
The gDoF outer bound coincides with that in FD~\cite{cardoneJ1}. Thus, as in FD, $\mathsf{GAP}  \leq  1$~bit.

\subsection{$S$-channel}
\label{sec:constGaps}
Let $\mathsf{I}_\mathsf{p} = \mathsf{I}$, $\mathsf{I}_\mathsf{c} = 0$ and $\mathsf{d}^{\rm(S)}=\mathsf{d}$ for brevity. We analyze different regimes.

{\bf{Regime 1: Very Strong Interference: $\alpha  \geq  2$}}. 
The gDoF is as for the non-cooperative ${S}$-channel.
In this regime the gDoF upper bound coincides with that in FD \cite{cardoneJ1}, so
$\mathsf{GAP}  =0 $~bit.

{\bf{Regime 2: $\alpha < 2$ and $\beta \leq \max \left \{ 1,\alpha \right \}$}}.  
In this regime the gDoF is as for the non-cooperative ${S}$-channel.
The gDoF upper bound coincides with that in FD \cite{cardoneJ1}. Thus, as in FD,
$\mathsf{GAP} \leq 2 \ \rm{bits}$.

{\bf{Regime 3: Weak Interference 1: $\alpha < 1, \ 1<\beta \leq 2-\alpha$.}}
\textit{Parameter Range}: $\mathsf{I} < \mathsf{S}$ and $\mathsf{S}<\mathsf{C}\leq \frac{\mathsf{S}^2}{\mathsf{I}}$, in which case we have $\mathsf{d}=1-\frac{\alpha}{2}$ (with $\gamma=0$). 
\textit{Inner Bound}: classical non-cooperative $S$-channel in weak interference~\cite[Theorem 2]{sason}, i.e., 
$\left({R}_\mathsf{p}+{R}_\mathsf{c} \right)^{\rm(IB)} = \log \left( 1+\mathsf{S} \right) + \log \left(  1+\frac{\mathsf{S}}{1+\mathsf{I}} \right)$,
which implies
$\mathsf{d} = 1-\frac{1}{2}\alpha$. 
\textit{Outer Bound}: in this regime the tightest sum-rate outer bound is given by \eqref{eq:Tuninetti}, that can be further bounded as
$\left({R}_\mathsf{p}+{R}_\mathsf{c} \right)^{\rm(OB)} \leq 2 + \log \left( 1+\left( \sqrt{\mathsf{I}}+\sqrt{\mathsf{S}} \right)^2 \right)+ \log \left( \frac{\mathsf{S}}{\mathsf{I}} \right)$.
Thus $\mathsf{GAP} \leq 2 + \log \left( 1+\frac{2\sqrt{\mathsf{S}\mathsf{I}}}{1+\mathsf{I}+\mathsf{S}}\right)+ \log \left( 1+\frac{1}{\mathsf{I}}\right) \leq 4 \ \rm{bits}$.

{\bf{Regime 4: Strong Interference: $1 \leq \alpha < 2, \ \beta > \alpha $.}}
\textit{Parameter Range}: $\mathsf{S} \leq \mathsf{I}<\mathsf{S}(\mathsf{S}+1)$ and $\mathsf{C} > \mathsf{I}$, in which case we have $2\mathsf{d} \leq \max_{\gamma} \min \left \{ \gamma+2\left(1-\gamma \right), \gamma \beta  + \left(1-\gamma \right) \alpha \right \}.$
In this expression the first term is decreasing in $\gamma$ while the second term is increasing in $\gamma$. Thus the optimal $\gamma$ can be found by equating the two terms and is given by $\gamma^\star = \frac{2-\alpha}{\beta-\alpha+1}$, which leads to
$\mathsf{d} \leq \frac{1}{2} \frac{2\beta-\alpha}{\beta -\alpha+1}$.
\textit{Inner Bound}: inspired by the LDA scheme in Figs. \ref{fig:firstphase} and \ref{fig:secondphase4S} (which can be inferred from that in Fig.\ref{fig:secondphase8} with different power splits),  
the rate in \eqref{eq:ach S regime 4} at the top of the page, with 
$\gamma^\prime = \frac{x}{\log \left( 1+\frac{\mathsf{C}}{1+\mathsf{S}} \right) + x}  \stackrel{\mathsf{SNR} \gg 1}{\to}  \gamma^\star$ for $x:= \log \left( 1+\frac{\mathsf{S}^2}{1+\mathsf{I}} \right)   \stackrel{\mathsf{SNR} \gg 1}{\to} 2-\alpha $, is achievable
The sum-rate in~\eqref{eq:ach S regime 4} implies
$2 \mathsf{d} 
 = 1 
  + \frac{2-\alpha}{\beta-\alpha+1} [-0+(\beta-1)]
  + \frac{\beta-1 }{\beta-\alpha+1} [\alpha-1]
  = \frac{2\beta-\alpha}{\beta -\alpha+1}$.
\textit{Outer Bound}: the tightest sum-rate outer bound is that in \eqref{eq:Tuninetti} (with $\mathsf{I}_\mathsf{c} = 0$).
With the first constraint in \eqref{eq:Tuninetti},
$\mathsf{GAP} \leq  4 \ \rm{bits}$;
with the second one in \eqref{eq:Tuninetti}, $\mathsf{GAP}  \leq  4 \ \rm{bits}$.
Thus, $\mathsf{GAP}  \leq   4 \ \rm{bits}$.

{\bf{Regime 5: Weak Interference 2: $\alpha < 1, \ \beta > 2-\alpha$}}.
\textit{Parameter Range}: $\mathsf{I} < \mathsf{S}$ and $\mathsf{C} > \frac{\mathsf{S}^2}{\mathsf{I}}$, in which case we have
$2\mathsf{d} \leq \max_{\gamma} \min \left \{ \gamma+2\left(1-\gamma \right),  \gamma \beta  + \left(1-\gamma \right) \left( 2-\alpha \right) \right \}.$
In this expression the first term is decreasing in $\gamma$ while the second term is increasing in $\gamma$. Thus the optimal $\gamma$ can be found by equating the two terms and is given by $\gamma^\star = \frac{\alpha}{\beta+\alpha-1}$,
which leads to
$\mathsf{d} \leq \frac{1}{2} \frac{2\beta+\alpha-2}{\beta +\alpha-1}$.
\textit{Inner Bound}: inspired by the LDA scheme in Figs. \ref{fig:firstphase} and \ref{fig:secondphase5S} (which can be inferred from that in Fig.\ref{fig:secondphase10} with different power splits),
the rate in \eqref{eq:ach S regime 5} at the top of the previous page, with 
$\gamma^\prime = \frac{x}{\log \left( 1+\frac{\mathsf{C}}{1+\mathsf{S}} \right)+x}  \stackrel{\mathsf{SNR} \gg 1}{\to} \gamma^\star$, with $x:= \log \left( 1 + \frac{\mathsf{S}\mathsf{I}}{1 + \mathsf{S}+\mathsf{I}} \right)  \stackrel{\mathsf{SNR} \gg 1}{\to} \min\{1,\alpha\}= \alpha$, is achievable.
The sum-rate in~\eqref{eq:ach S regime 5} implies
$2\mathsf{d} 
= \frac{\alpha }{\beta+\alpha-1}[1-0+(\beta-1)]
+ \frac{\beta-1}{\beta+\alpha-1}[(1-\alpha)+1-0]
= \frac{2\beta+\alpha-2}{\beta +\alpha-1}$.
\textit{Outer Bound}: the tightest sum-rate outer bound is that in \eqref{eq:Tuninetti} (with $\mathsf{I}_\mathsf{c} = 0$).
With the first constraint in \eqref{eq:Tuninetti},
$\mathsf{GAP} \leq  4 \ \rm{bits}$;
with the second one in \eqref{eq:Tuninetti}, $\mathsf{GAP}  \leq  5 \ \rm{bits}$.
Thus, $\mathsf{GAP}  \leq   5 \ \rm{bits}$.

\section{}
\label{app:ach}
Here we develop achievable schemes inspired by Fig.~\ref{fig:fig5}.
In the following all signals $\mathsf{X}_{b_j}$ for some subscript $j$, are independent proper-complex Gaussian random variables with zero mean and unit variance and represent codebooks used to convey the bits in $b_j$ in Fig.~\ref{fig:fig5}.

\subsection{Phase~1 of duration $\gamma\in[0,1]$ (see also Fig.~\ref{fig:firstphase})}
\label{app:first phase}
The transmitted signals are
$\mathsf{X}_\mathsf{p}[1]=\sqrt{1-\eta} \mathsf{X}_{b_{1}} + \sqrt{\eta} \mathsf{X}_{b_{2}}$, with $\eta := \frac{1}{1+\mathsf{S}}$ and $\mathsf{X}_\mathsf{c}[1]=0$.
CTx applies successive decoding of $\mathsf{X}_{b_{1}}$ followed by $\mathsf{X}_{b_{2}}$ from $\mathsf{Y}_\mathsf{f}[1]$
which is possible if 
\begin{align}
&R_{b_{1}} \leq \gamma \log \left( 1+\mathsf{C} \right) - \gamma \log \left( 1+\frac{\mathsf{C}}{1+\mathsf{S}} \right ) \nonumber
\\& R_{b_{2}} \leq \gamma \log \left ( 1+\frac{\mathsf{C}}{1+\mathsf{S}} \right ).
\label{eq:eq1}
\end{align}
PRx decodes $\mathsf{X}_{b_{1}}$ treating $\mathsf{X}_{b_{2}}$ as noise from $\mathsf{Y}_\mathsf{p}[1]$
which is possible if 
\begin{align}
&R_{b_{1}} \leq \gamma \log \left( 1+\mathsf{S}\right) - \gamma \log \left( 1+\frac{\mathsf{S}}{1+\mathsf{S}} \right).
\label{eq:eq2}
\end{align}
Since $\mathsf{C}>\mathsf{S}$, Phase 1 is successful if \eqref{eq:eq1} and \eqref{eq:eq2} are satisfied.

\subsection{Phase~2 of duration $\left( 1-\gamma \right)$ for Region 3 in Fig. \ref{fig:fig3} (see also Fig.~\ref{fig:secondphase3})}
\label{sec:ach3sym}
The transmit signals are
$\mathsf{X}_\mathsf{p}[2]= \mathsf{X}_{b_{3c}}$ and $\mathsf{X}_\mathsf{c}[2]= \sqrt{\eta} \mathsf{X}_{b_{2}}+ \sqrt{1-\eta} \mathsf{X}_{b_{4c}}$, with $\eta = \frac{1}{1+\mathsf{S}}$.
PRx applies successive decoding as follows: $\mathsf{X}_{b_{4c}}$, $\mathsf{X}_{b_{2}}$, $\mathsf{X}_{b_{3c}}$ from $\mathsf{Y}_\mathsf{p}[2]$,
which is possible if 
\begin{align}
  & R_{b_{4c}}\leq \left( 1\!-\!\gamma \right) \left(\log \left( 1\!+\!\mathsf{S}\!+\!\mathsf{I} \right) \!-\! \log \left( 1\!+\!\frac{\mathsf{I}}{1\!+\!\mathsf{S}}\!+\!\mathsf{S} \right) \right) \label{eq:eq2something}
\\& R_{b_2} \leq \left( 1-\gamma \right) \log \left( 1+\frac{\mathsf{I}}{(1+\mathsf{S})^2} \right) \label{eq:eq3}
\\& R_{b_{3c}} \leq \left( 1-\gamma \right)\log\left(1+\mathsf{S}\right)\label{eq:eq4}.
\end{align}
CRx successively decodes $\mathsf{X}_{b_{3c}}$ and $\mathsf{X}_{b_{4c}}$ (treating $\mathsf{X}_{b_{2}}$ as noise) from $\mathsf{Y}_\mathsf{c}[2]$,
which is possible if 
\begin{align}
&R_{b_{3c}} \leq \left( 1-\gamma \right) \log \left( 1+\mathsf{S}+\mathsf{I} \right)-\left( 1-\gamma \right) \log \left( 1+\mathsf{S}\right)\label{eq:eq5something}
\\& R_{b_{4c}} \leq  \left( 1-\gamma \right) \log \left( 1+\mathsf{S}\right) - \left( 1-\gamma \right) \log \left( 1\!+\!\frac{\mathsf{S}}{1\!+\!\mathsf{S}}\right). \label{eq:eq6}
\end{align}
Phase~2 is successful if
$\min  \{\text{eq.\eqref{eq:eq4},eq.\eqref{eq:eq5something}\} = eq.\eqref{eq:eq4}}$,
$\min  \{\text{eq.\eqref{eq:eq6},eq.\eqref{eq:eq2something}\} = eq.\eqref{eq:eq6}}$
and
eq.\eqref{eq:eq3}, are satisfied.
By imposing that $R_{b_{2}}$ is the same in both phases, i.e., that \eqref{eq:eq1} and \eqref{eq:eq3} are equal, we get that $\gamma$ should be chosen equal to 
$ \gamma^\prime = \frac{x}{\log \left( 1+\frac{\mathsf{C}}{1+\mathsf{S}} \right)+x}, \quad x := \log \left( 1+\frac{\mathsf{I}}{(1+\mathsf{S})^2} \right)$.
Therefore the total sum-rate decoded at PRx and CRx through the two phases is
$\left({R}_\mathsf{p}+{R}_\mathsf{c} \right)^{\rm(IB)} = R_{b_{1}}+R_{b_{2}}+R_{b_{3c}}+R_{b_{4c}}$ as given in~\eqref{eq:ach1a}.

\subsection{Phase~2 of duration $\left( 1-\gamma \right)$ for Region 8 in Fig. \ref{fig:fig3} (see also Fig.~\ref{fig:secondphase8})}
\label{sec:ach8sym}
The transmitted signals are
\begin{align*}
& \mathsf{X}_\mathsf{p}[2] =  \sqrt{\delta_1} \mathsf{X}_{b_{3c}} + \sqrt{\delta_2} \mathsf{X}_{b_2} + \sqrt{\delta_3} \mathsf{X}_{b_{3p}}
\\& \mathsf{X}_\mathsf{c}[2] =  \sqrt{\delta_3} \mathsf{X}_{b_{4p}} +  \sqrt{1-\delta_3} \mathsf{X}_{b_{4c}}
\end{align*}
with $ \delta_1 = 1- \delta_2-\delta_3, 
\quad \delta_2 =  \frac{\mathsf{S}}{(1 + \mathsf{I})^2}, 
\quad \delta_3 = \frac{1}{1 + \mathsf{I}}$,
where $\mathsf{X}_{b_{4p}}$ is DPC-ed against $\mathsf{X}_{b_2}$ at $\mathsf{Y}_\mathsf{c}[2]$.
PRx applies successive decoding as follows: $\mathsf{X}_{b_{3c}}$, $\mathsf{X}_{b_{2}}$, $\mathsf{X}_{b_{4c}}$ and $\mathsf{X}_{b_{3p}}$ (treating $\mathsf{X}_{b_{4p}}$ as noise) from $\mathsf{Y}_\mathsf{p}[2]$,
which is possible if 
\begin{align}
&R_{b_{3c}} \leq \left( 1-\gamma \right) \log \left( 1+\mathsf{S}+\mathsf{I} \right) \nonumber
\\& \qquad -\left( 1-\gamma \right) \log \left( 1+\mathsf{I}+\frac{\mathsf{S}^2+\mathsf{S}\mathsf{I}+\mathsf{S}}{(1+\mathsf{I})^2} \right) \nonumber
\\& R_{b_2} \leq \left( 1-\gamma \right) \log \left( 1+\mathsf{I}+\frac{\mathsf{S}^2+\mathsf{S}\mathsf{I}+\mathsf{S}}{(1+\mathsf{I})^2} \right) \nonumber
\\& \qquad -\left( 1-\gamma \right) \log \left( 1+\mathsf{I}+\frac{\mathsf{S}}{1+\mathsf{I}}\right) \label{eq:eq3aa}
\\ & R_{b_{4c}} \leq \left( 1-\gamma \right) \log \left( 1+\mathsf{I}+\frac{\mathsf{S}}{1+\mathsf{I}}\right) \nonumber
\\& \qquad - \left( 1-\gamma \right) \log \left( 1+\frac{\mathsf{S}}{1+\mathsf{I}}+\frac{\mathsf{I}}{1+\mathsf{I}} \right) \label{eq:eq4aa}
\\& R_{b_{3p}} \leq \left( 1-\gamma \right) \log \left( 1+\frac{\mathsf{S}}{1+\mathsf{I}}+\frac{\mathsf{I}}{1+\mathsf{I}} \right)\nonumber
\\& \qquad -\left( 1-\gamma \right) \log \left( 1+\frac{\mathsf{I}}{1+\mathsf{I}} \right). \label{eq:eq5aa}
\end{align}
CRx applies successive decoding as follows: $\mathsf{X}_{b_{4c}}$, $\mathsf{X}_{b_{3c}}$ and $\mathsf{X}_{b_{4p}}$ from $\mathsf{Y}_\mathsf{c}[2]$, which is possible if 
\begin{align}
&R_{b_{4c}} \leq \left( 1-\gamma \right) \left (\log \left( 1+\mathsf{S}+\mathsf{I} \right)- \log \left( 1+\mathsf{I}+\frac{\mathsf{S}}{1+\mathsf{I}} \right) \right ) \nonumber
\\& R_{b_{3c}} \leq \left( 1-\gamma \right) \log \left( 1+\mathsf{I}+\frac{\mathsf{S}}{1+\mathsf{I}} \right) \nonumber
\\& \qquad -\left( 1-\gamma \right) \log \left( 1+ \frac{\mathsf{S}\mathsf{I}+\mathsf{I}+\mathsf{I}^2}{(1+\mathsf{I})^2} + \frac{\mathsf{S}}{1+\mathsf{I}}\right) \label{eq:eq6aa}
\\& R_{b_{4p}} \leq \left( 1-\gamma \right) \log \left( 1+\frac{\mathsf{S}}{1+\mathsf{I}}+\frac{\mathsf{I}}{1+\mathsf{I}} \right)\nonumber
\\& \qquad -\left( 1-\gamma \right) \log \left( 1+\frac{\mathsf{I}}{1+\mathsf{I}} \right).\label{eq:eq7aa}
\end{align}
Thus, since we are in the regime $\mathsf{I}(1+\mathsf{I})^2<\mathsf{S}(\mathsf{S}+1)$, Phase~2 is successful if \eqref{eq:eq3aa}, \eqref{eq:eq4aa}, \eqref{eq:eq5aa}, \eqref{eq:eq6aa} and \eqref{eq:eq7aa} are satisfied.  
By imposing that $R_{b_{2}}$ is the same in both phases, that is, that \eqref{eq:eq1} and \eqref{eq:eq3aa} are equal, we get that $\gamma$ should be chosen equal to 
 $\gamma^\prime = \frac{x}{\log \left( 1+\frac{\mathsf{C}}{1+\mathsf{S}} \right)+x}, \quad x := \log \left( 1+\frac{\mathsf{S}^2}{(1+\mathsf{I})^3+\mathsf{S}+\mathsf{S}\mathsf{I}} \right).$
With this scheme, the total sum-rate decoded at PRx and CRx through the two phases is
$\left({R}_\mathsf{p} + {R}_\mathsf{c}\right)^{\rm(IB)} =  R_{b_1}+R_{b_{3c}}+R_{b_{3p}}+R_{b_{2}}+R_{b_{4c}}+R_{b_{4p}}$ as given in~\eqref{eq:eqach3a}.

\subsection{Phase~2 of duration $\left( 1-\gamma \right)$ for Region 10 in Fig. \ref{fig:fig3} (see also Fig.~\ref{fig:secondphase10})}
\label{sec:ach10sym}

The transmitted signals are
$\mathsf{X}_\mathsf{p}[2]= \sqrt{1-\delta} \mathsf{X}_{b_{2}}+ \sqrt{\delta} \mathsf{X}_{b_{3p}}$ with $\delta = \frac{\mathsf{1}}{1+\mathsf{I}}$ and
$\mathsf{X}_\mathsf{c}[2] =   \mathsf{X}_{b_{4p}}$, which is DPC-ed against $\mathsf{X}_{b_{2}}$ at $\mathsf{Y}_\mathsf{c}[2]$.
PRx decodes $\mathsf{X}_{b_{2}}$ and $\mathsf{X}_{b_{3p}}$ in this order (by treating $\mathsf{X}_{b_{4p}}$ as noise) from $\mathsf{Y}_\mathsf{p}[2]$, that is possible if 
\begin{align}
& R_{b_2}    \leq \left( 1\!-\!\gamma \right) \left (\log \left( 1\!+\!\mathsf{S}\!+\!\mathsf{I} \right)
\!-\! \log \left( 1\!+\!\mathsf{I}\!+\!\frac{\mathsf{S}}{1+\mathsf{I}} \right) \right )\label{eq:eq3a}
\\& R_{b_{3p}} \leq \left( 1-\gamma \right) \left (\log \left( 1+\mathsf{I}+\frac{\mathsf{S}}{1+\mathsf{I}} \right)- \log \left(1+ \mathsf{I}\right)\!\right ) \label{eq:eq4a} .
\end{align}
CRx decodes $\mathsf{X}_{b_{4p}}$ (by treating $\mathsf{X}_{b_{3p}}$ as noise) from $\mathsf{Y}_\mathsf{c}[2]$, which is possible if 
\begin{align}
 &R_{b_{4p}}\!\leq\! \left( 1\!-\!\gamma \right) \left (\log \left( 1\!+\! \mathsf{S}\!+\!\frac{\mathsf{I}}{1\!+\!\mathsf{I}} \right) \!-\!  \log \left(1\!+\!\frac{\mathsf{I}}{1\!+\!\mathsf{I}}\right)\right).\label{eq:eq6a}
\end{align}
Thus Phase~2 is successful if \eqref{eq:eq3a}, \eqref{eq:eq4a} and \eqref{eq:eq6a} are satisfied.  
By imposing that $R_{b_{2}}$ is the same in both phases, that is, that \eqref{eq:eq1} and \eqref{eq:eq3a} are equal, we get that $\gamma$ should be chosen equal to 
 $\gamma^\prime = \frac{x}{\log \left( 1+\frac{\mathsf{C}}{1+\mathsf{S}} \right)+x}, \quad x := \log \left( 1+\frac{\mathsf{S}\mathsf{I}}{(1+\mathsf{I})^2+\mathsf{S}} \right).$
With this scheme, the total sum-rate decoded at PRx and CRx through the two phases is
$\left({R}_\mathsf{p} + {R}_\mathsf{c}\right)^{\rm(IB)} = R_{b_{1}}+R_{b_{3p}}+R_{b_{2}}+R_{b_{4p}}$ as given in~\eqref{eq:eqach2a}.

\bibliographystyle{IEEEtran}
\bibliography{JsacBib}

\end{document}